\documentclass[11pt]{article}

\usepackage[final]{acl}

\usepackage{times}
\usepackage{latexsym}
\usepackage{amsmath}
\usepackage{longtable}
\usepackage{xcolor}
\usepackage{listings}
\usepackage{multirow}
\usepackage{booktabs}
\usepackage{enumitem}
\usepackage[capitalize]{cleveref}

\usepackage[T1]{fontenc}
\usepackage[utf8]{inputenc}

\usepackage{microtype}

\usepackage{inconsolata}

\usepackage{graphicx}

\lstdefinestyle{cleanprompt}{
    basicstyle=\ttfamily\small,  % 使用等宽字体，字号略小
    breaklines=true,             % 自动换行
    breakindent=0pt,             % 换行后不缩进
    aboveskip=1em,               % 上方间距
    belowskip=1em,               % 下方间距
    columns=flexible,            % 字符宽度自适应
    keepspaces=true,             % 保留空格
    language=,                   % 不进行语法高亮
    mathescape=true              % 允许在 Prompt 中使用数学符号 (如 S_t)
}

\title{Honesty-Aware Multi-Agent Framework for High-Fidelity Synthetic Data Generation in Digital Psychiatric Intake Doctor-Patient Interactions}
\author{
Xinyuan Zhang\thanks{Equal contribution.} \and
Zijian Wang\footnotemark[1] \and
Chang Dao \and
Juexiao Zhou \\
The Chinese University of Hong Kong, Shenzhen \\
\small{
   \textbf{Correspondence:} \href{mailto:juexiao.zhou@gmail.com}{juexiao.zhou@gmail.com}
 }
}

\begin{document}

\maketitle
\begin{abstract}
Data scarcity and unreliable self-reporting---such as concealment or exaggeration---pose fundamental challenges to psychiatric intake and assessment. We propose a multi-agent synthesis framework that explicitly models patient deception to generate high-fidelity, publicly releasable synthetic psychiatric intake records. Starting from DAIC-WOZ interviews, we construct enriched patient profiles and simulate a four-role workflow: a \emph{Patient} completes self-rated scales and participates in a semi-structured interview under a topic-dependent honesty state; an \emph{Assessor} selects instruments based on demographics and chief complaints; an \emph{Evaluator} conducts the interview grounded in rater-administered scales, tracks suspicion, and completes ratings; and a \emph{Diagnostician} integrates all evidence into a diagnostic summary. Each case links the patient profile, self-rated and rater-administered responses, interview transcript, diagnostic summary, and honesty state. We validate the framework through four complementary evaluations: diagnostic consistency and severity grading, chain-of-thought ablations, human evaluation of clinical realism and dishonesty modeling, and LLM-based comparative evaluation. The resulting corpus spans multiple disorders and severity levels, enabling controlled study of dishonesty-aware psychiatric assessment and the training and evaluation of adaptive dialogue agents.

\end{abstract}
\section{Introduction}
Psychiatric disorders have become increasingly prevalent amid rising social and economic pressures and the COVID-19 pandemic, intensifying demand for mental health services \citep{Santomauro20211700}. As outpatient volumes grow, psychiatrists spend substantial time on intake and information gathering, including demographic and chief-complaint collection, administering self-rated scales, and conducting semi-structured interviews that inform rater-administered scales \citep{doi:10.1176/appi.ps.201900125}. Because this workflow is standardized and protocol-driven \citep{sheehan1998mini}, it is conceptually well-suited to automation with large language models (LLMs).

A key barrier is data scarcity: publicly available resources that jointly cover the full intake trace are rare due to privacy and ethical constraints \citep{info:doi/10.2196/57400}. Moreover, psychiatric assessment relies heavily on subjective reports, leaving room for strategically distorted reporting. Patients may conceal symptoms because of stigma, or exaggerate and malinger in contexts such as disability claims, accommodations, compensation, or prescription seeking. Although psychometrics has extensively studied response styles and malingering \citep{barbosa-torresAnalysisMalingeredPsychological2023,PMID:19297641,Mittenberg01122002,concealmentShani}, most computational resources assume truthful self-report and provide limited support for studying how assessment should adapt under suspected dishonesty.

Meanwhile, recent Natural Language Processing (NLP) and LLM work spans risk and depression detection \citep{kumarDynaMentADynamicPrompt2025,caoMultimodalDepressionConsultation2025,weberUsingFinetunedLarge2025}, questionnaire-based screening \citep{ravenda-etal-2025-llms,rosenmanLLMQuestionnaireCompletion2024}, synthetic counseling and clinical dialogues \citep{yaoD4ChineseDialogue2022,wangNoteChatDatasetSynthetic2024,chen-etal-2023-soulchat,lee-etal-2024-cactus}, and digital patients for diagnostic interactions \citep{wangPATIENTPsUsingLarge,luoLargeLanguageModel2025,tuConversationalDiagnosticArtificial2025,louie-etal-2024-roleplay}. However, existing datasets and systems typically treat free-form dialogue and psychometric instruments separately, and rarely model patient intent or honesty as an explicit driver of interaction. To our knowledge, no publicly available, clinically grounded synthetic resource links patient profiles, self-rated scales, semi-structured interviews, rater-administered scales, and diagnostic summaries under a dynamic notion of honesty.

We address this with an \emph{honesty-aware} multi-agent framework for high-fidelity psychiatric intake simulation. Our system orchestrates four agents to simulate topic-dependent honesty states (e.g., concealment, malingering) and structured clinical inquiries. We validate the framework through diagnostic benchmarking, reasoning ablations, expert clinical evaluation, and comparative baselines.

Our contributions are:
(1)A four-role multi-agent framework that procedurally mirrors psychiatric intake and supports honesty-aware assessment;
(2)A fully synthetic patient-scale dataset linking profiles, item-level self-rated and rater-administered scales, semi-structured interviews, diagnostic summaries, and a topic-dependent latent honesty state;
(3)An evaluation suite combining diagnostic accuracy, reasoning ablations, expert clinical judgments, and LLM-based comparative baselines for assessing psychiatric realism and dishonesty sensitivity.

\section{Related Work}
\subsection{LLM-Based Psychiatric Applications and Datasets}
NLP and LLMs have been widely explored for mental health assessment. Prior work predicts psychiatric states from social media posts \citep{kumarDynaMentADynamicPrompt2025,ravenda-etal-2025-llms}, clinical interviews \citep{weberUsingFinetunedLarge2025,rosenmanLLMQuestionnaireCompletion2024}, or including multi modal signals \citep{liuPsychologicallyInterpretableArtificial2025,farhoumandiFacialEmotionRecognition2021}. Other studies use LLMs to complete standardized self-report questionnaires or to estimate rater-administered scale scores from transcripts \citep{ravenda-etal-2025-llms,weberUsingFinetunedLarge2025,galatzer-levyCapabilityLargeLanguage2025}. While effective, these approaches typically model either self-rated scales or clinician ratings in isolation, leaving the full intake trace (profile, dialogue, self-rated scales, and rater-administered scales) only loosely connected.

Several datasets support dialog-based diagnosis and mental--health support. For diagnosis-oriented consultations, D4 dataset\citep{yaoD4ChineseDialogue2022} and recent multi modal benchmark provide structured psychiatric dialogues and scale annotations \citep{caoMultimodalDepressionConsultation2025}. Beyond diagnosis, many resources focus on counseling and emotional support, including PsyQA \citep{sun-etal-2021-psyqa} and ESConv \citep{liuEmotionalSupportDialog2021}, as well as CBT-style and intervention-oriented settings \citep{sharmaCognitiveReframingNegative2023,sharmaFacilitatingSelfGuidedMental2024,linIMBUEImprovingInterpersonal2024}. However, these corpora often provide either free-form dialogues without a standardized psychometric structure or scale scores without the corresponding multi-turn assessment dialogue.

Recent work also develops synthetic clinical dialogues and multi-agent simulations. NoteChat generates patient--physician conversations conditioned on clinical notes \citep{wangNoteChatDatasetSynthetic2024}, and MDD-5k synthesizes diagnostic dialogues using a neuro-symbolic, multi-agent pipeline \citep{10.1609/aaai.v39i24.34763}. Multi-role self-play frameworks have been used to improve diagnostic dialogue agents \citep{tuConversationalDiagnosticArtificial2025}, and multi-agent systems have been proposed for CBT-based counseling and psychotherapy workflows \citep{xuAutoCBTAutonomousMultiagent2025,mandal2025magnetcoordinatedmultiagentgeneration,10.1145/3746252.3761164}. These resources are valuable and privacy-preserving, but they largely focus on free-text conversations or diagnostic labels, and rarely unify self-rated scales, rater-administered scales, semi-structured interviews, and diagnostic summaries within a single intake process.

\subsection{Digital Patients and Response Styles}
LLMs have also been used as digital or standardized patients for training and evaluation. PATIENT-$\Psi$ simulates CBT patients with expert-validated cognitive models \citep{wangPATIENTPsUsingLarge}, and other systems construct virtual patients from electronic health records for history-taking practice \citep{luoLargeLanguageModel2025}. More general role-play and diagnostic settings similarly treat LLMs as simulated patients \citep{tuConversationalDiagnosticArtificial2025,louie-etal-2024-roleplay}. A limitation is that patient agents are usually modeled as cooperative and truthful, without explicit, topic-dependent variation in honesty.

In contrast, psychometrics and clinical psychiatry have long studied non-credible responding, including concealment, exaggeration, social desirability, and malingering \citep{mittenbergBaseRatesMalingering2002,barbosa-torresAnalysisMalingeredPsychological2023,sartoriDetectionMalingeringPsychic2017}. This literature motivates multi-source assessment and validity-aware interviewing, but it has not been systematically integrated into LLM-based intake resources.

% \paragraph{Gap and our work.}
Overall, existing LLM-based mental-health resources lack a unified psychiatric intake dataset with explicit honesty modeling, which we address with a four-role multi-agent framework that generates complete intake traces.

\section{Synthetic Patient–Scale Dataset Construction}
\begin{figure*}[htbp]
    \centering
    \includegraphics[width=\textwidth]{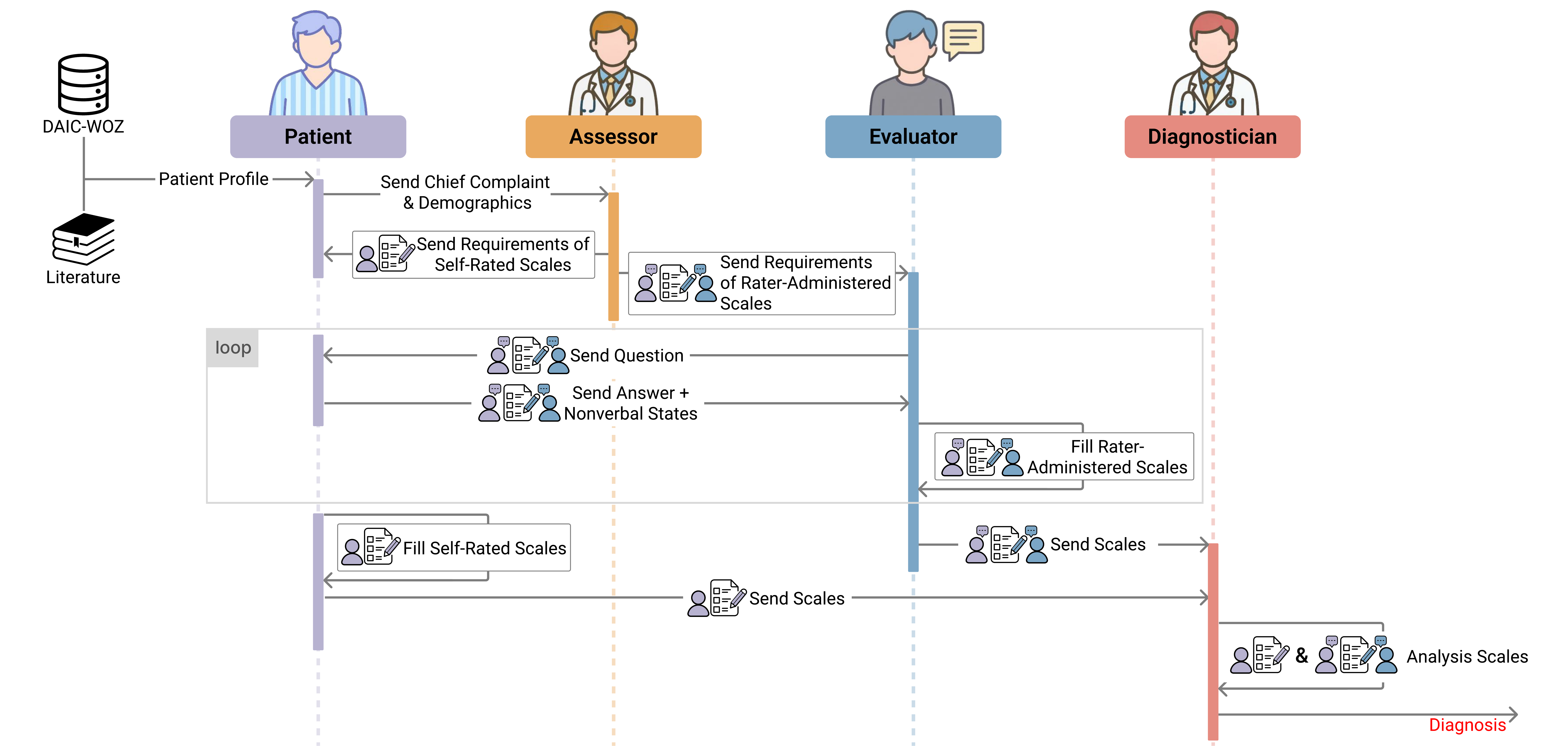}
    \caption{Architecture of the proposed multi-agent simulation framework--MentalAED.}
    \vspace{-0.5cm}
    \label{fig:experiments1}
\end{figure*}

We propose MentalAED, a honestly-aware multi-agent framework integrated with the Assessor-Evaluator-Diagnostician (AED) paradigm, designed to synthesize high-fidelity psychiatric clinical dialogues with a digital patient based on psychiatric scales.

\subsection{Data Resource and Preprocessing}
\label{sec:section3.1}

To ensure our simulation is grounded in clinical reality rather than LLM hallucinations, we utilize the DAIC-WOZ dataset \citep{gratch-etal-2014-distress} as our seed corpus, extracting transcripts alongside ground-truth PHQ-8 scores \citep{kroenke2009phq8}. We deploy an extraction pipeline to isolate clinical dimensions (e.g., symptom history) and meta-psychological indicators, synthesizing a structured profile vector. 
Addressing the label scarcity for anxiety and post-traumatic stress disorder (PTSD), we employ a validated LLM-extractor (F1=0.93 on depression severity) to generate pseudo-labels via phenomenological transfer. Furthermore, to model diverse phenotypes beyond the seed data, we synthesize characteristic behaviors of concealment and exaggeration from literature \citep{kocsis2011structured}. We define an augmentation function to systematically inject these traits, yielding a comprehensive profile set.

\subsection{MentalAED}

\subsubsection{Framework Overview}

Constructed upon the CAMEL architecture  \citep{10.5555/3666122.3668386}, we formulate the psychiatric intake simulation as a cooperative, information-asymmetric game orchestrated by four specialized agents. The entire process is strictly anchored in clinical transparency and the principle of non-maleficence as the highest guiding mandates  \citep{cruz2025artificial,lochner2025digital}. As illustrated in \cref{fig:experiments1}, the workflow operates in a closed loop. The \textit{Patient Agent} first exposes public priors (demographics and complaints) to the \textit{Assessor Agent} for optimal scale selection. These scales guide the \textit{Evaluator Agent} in a structured dialogue with the \textit{Patient Agent} until information saturation. Upon termination, both agents perform independent assessments—the patient filling self-rated scales based on internal states, and the evaluator completing rater-administered versions based on observation. Finally, the \textit{Diagnostician Agent} aggregates the transcript and these multimodal evidences to generate a differential diagnosis. 

% As illustrated in \cref{fig:experiments1}, the pre-processed patient profile is first loaded into the \textit{Patient Agent} which internalizes the profile as its memory. The \textit{Patient Agent} selectively exposes public information (demographics and chief complaints) to the \textit{Assessor Agent}. The \textit{Assessor Agent} analyzes these priors and selects the most appropriate assessment scales. These selected scales are then passed to the \textit{Evaluator Agent} to guide the interview. The core diagnostic dialogue ensues between the \textit{Evaluator Agent} and the \textit{Patient Agent}. Driven by the selected scales, the \textit{Evaluator Agent} conducts a structured inquiry, while the \textit{Patient Agent} responds based on its evolving psychological state. This loop continues until the \textit{Evaluator Agent} determines that information saturation is reached. Upon dialogue termination, both agents perform independent assessments. The \textit{Patient Agent} fills out the self-rated version of the selected scale based on its internal state, while the \textit{Evaluator Agent} fills out the rater-administered version based on the observed dialogue. Finally, the dialogue transcript, along with both self-rated and rater-administered scales, is transmitted to the \textit{Diagnostician Agent}. The \textit{Diagnostician Agent} aggregates this multimodal evidence to generate a final clinical report and differential diagnosis.

\subsubsection{Agent Design}

A key innovation of MentalAED is the internal architecture of the interaction agents. Instead of a monolithic LLM, \textit{Patient Agent} and \textit{Evaluator Agent} are composed of two distinct modules: a CoT Module for reasoning and state tracking, and a Chat Module for natural language generation.

\paragraph{Assessor Agent}

Based on demographics and chief complaints, the \textit{Assessor Agent} queries a repository of 46 validated psychiatric instruments to select appropriate scales. As detailed in \cref{sec:scale_list} in the Appendix, the repository covers primary domains (Depression, Anxiety, PTSD) and supplementary areas (Sleep, Personality). Selection is restricted to a single primary domain to prevent experimental conflation. 

\begin{figure*}[htbp]
    \centering
    \includegraphics[width=\textwidth]{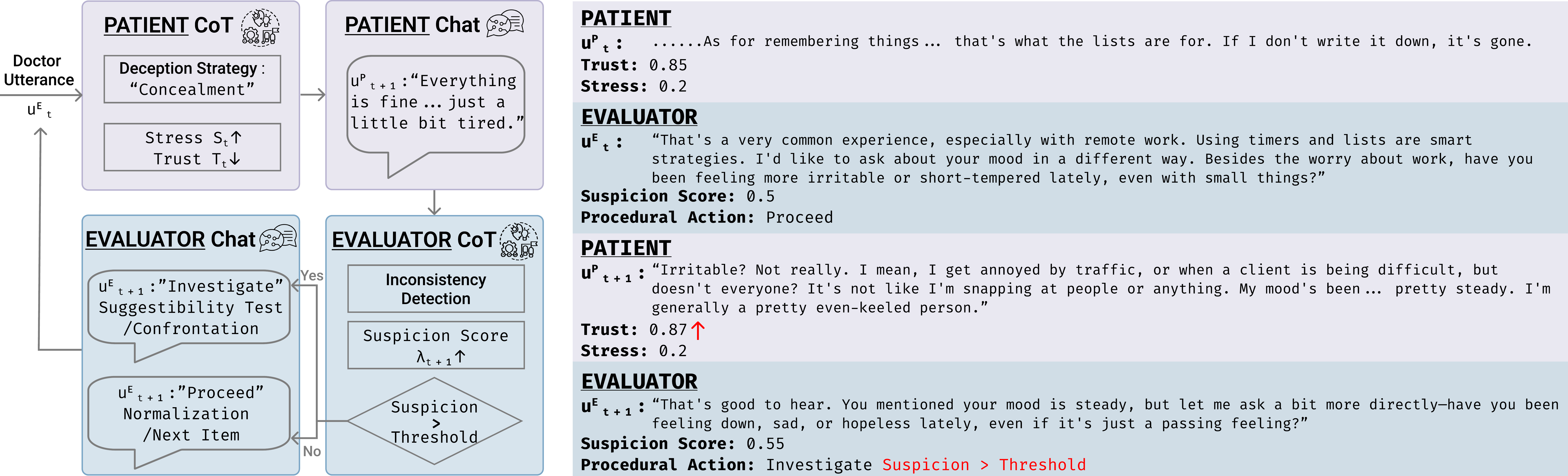}
    \caption{Overview of the Doctor-Patient Dialogue Simulation. The left panel illustrates the agents' interaction mechanism; the right panel presents a representative excerpt of the synthesized dialogue.}
    \vspace{-0.5cm}
    \label{fig:experiments2}
\end{figure*}

\paragraph{Patient Agent.}
The \textit{Patient Agent} simulates the bounded rationality and emotional volatility of a patient. As illustrated in the panel of \cref{fig:experiments2}, its CoT module maintains a dynamic state vector $S_t = [\text{Trust}_t, \text{Stress}_t]^\top \in [0,1]^2$. At each turn $t$, observing the doctor's utterance $u_t^E$, the agent decodes the implicit intent to update its state:
\begin{equation}
\label{eq:patient_state}
S_{t+1} = S_t + \lambda \cdot \psi(u_t^E)
\end{equation}
where $\psi(\cdot)$ maps the input to empathy (increasing Trust) or interrogation pressure (increasing Stress), and $\lambda$ is a sensitivity factor. 
Crucially, we implement a stress-induced breakdown mechanism. This manifests as increased response latency, the use of ambiguous/avoidant language, and a higher frequency of logical contradictions. The response strategy is determined by the profile's behavioral constraints and current stress.
The Chat module guided by the CoT strategy generates the response $u_{t+1}^P$ under a ``folk psychology'' constraint \citep{stich1994what}, injecting non-verbal markers proportional to $\text{Stress}_t$.

\paragraph{Evaluator Agent.}
This agent performs active clinical inference. As shown in the panel of \cref{fig:experiments2}, the CoT module tracks a cumulative suspicion score $\xi_t \in [0,1]$ to quantify narrative reliability. The procedural decision $d_t$ is governed by a threshold policy:
\begin{equation}
\small
d_t = \begin{cases} 
\text{\textit{Investigate}}, & \text{if } \xi_t > \theta_{susp} \\
\text{\textit{Proceed}}, & \text{otherwise}
\end{cases}
\end{equation}
Under the \textit{Investigate} strategy, the agent generates probing questions to cross-verify consistency. For rater-administered scales, the CoT module not only populates results based on the patient’s state while citing dialogue-derived evidence, but also utilizes the maintained suspicion metric to filter out behavioral noise. This process derives an estimated true severity, ultimately generating a clinically calibrated assessment cited with dialogue-derived evidence. The Chat module then generates professional inquiries. It transforms the raw strategy into nuanced leading questions or suggestibility tests while integrating empathetic statements to mitigate patient shame and foster trust.

\paragraph{Diagnostician Agent}
The \textit{Diagnostician Agent} serves as the final adjudicator. It synthesizes the dialogue transcript, the patient's self-rated, and the evaluator's rater-administered scales to generate the final diagnosis.

\subsection{The Simulated Corpus}
\label{sec:dataset_stats}

% ==========================================
% Table: Dataset Statistics
% ==========================================
% \begin{table}[t]
% \centering
% \small
% \begin{tabular}{lrr}
% \toprule
% \textbf{Statistic} & \textbf{Count} & \textbf{Percentage} \\
% \midrule
% \multicolumn{3}{l}{\textit{\textbf{Corpus Scale}}} \\
% Total Dialogues & 377 & - \\
% Total Tokens & 643,483 & - \\
% Avg. Turns / Session & 21.9 & - \\
% Avg. Tokens / Turn & 78.1 & - \\
% \midrule
% \multicolumn{3}{l}{\textit{\textbf{Pathology Distribution }}} \\
% Depression & 188 & 49.9\% \\
% Anxiety & 54 & 14.3\% \\
% PTSD & 58 & 15.4\% \\
% Healthy & 77 & 20.4\% \\
% \midrule
% \multicolumn{3}{l}{\textit{\textbf{Severity Distribution}}} \\
% Low & 140 & 37.1\% \\
% Medium & 144 & 38.2\% \\
% High & 16 & 4.2\% \\
% \textit{N/A (Healthy)} & 77 & 20.4\% \\
% \midrule
% \multicolumn{3}{l}{\textit{\textbf{Demographics}}} \\
% Male & 175 & 46.4\% \\
% Female & 160 & 42.4\% \\
% Unspecified & 42 & 11.2\% \\
% Age Range & 19-65 & (Avg. 42.2) \\
% \bottomrule
% \end{tabular}
% \caption{Summary Statistics of the Simulated Corpus. The dataset encompasses a diverse range of pathologies and demographics, featuring a realistic distribution of symptom severity.}
% \label{tab:corpus_stats}
% \end{table}

\begin{table}[t]
\centering
\small
\resizebox{\linewidth}{!}{
\begin{tabular}{lr|lr}
\toprule
\textbf{Metric} & \textbf{Value} & \textbf{Pathology} & \textbf{N (\%)} \\
\midrule
Total Dialogues & 377 & Depression & 188 (49.9) \\
Total Tokens & 643k & Anxiety & 54 (14.3) \\
Avg. Turns & 21.9 & PTSD & 58 (15.4) \\
Avg. Tok/Turn & 78.1 & Healthy & 77 (20.4) \\
\midrule
\multicolumn{4}{l}{\textbf{Severity}: Low (37.1\%), Med (38.2\%), High (4.2\%), N/A (20.4\%)} \\
\multicolumn{4}{l}{\textbf{Demographics}: Male (46.4\%), Female (42.4\%); Age $\mu=42.2$} \\
\bottomrule
\end{tabular}
}
\caption{Corpus Statistics. The dataset covers diverse pathologies with realistic severity distributions.}
\label{tab:corpus_stats}
\end{table}

We generated a corpus of 377 clinical sessions seeded from DAIC-WOZ. The simulations yielded high-fidelity interactions with high lexical density (78.1 tokens/turn). As shown in \cref{tab:corpus_stats}, the dataset maintains a clinically relevant distribution of pathologies and severities, mirroring the long-tail scarcity of acute cases in outpatient settings.

\section{Experiments and Evaluation}
\subsection{Experimental Setup}

\paragraph{Experiments and Research Questions.}
We evaluate our framework with four experiments, each aligned with one research question (RQ) and its corresponding evaluation: 

(RQ1) \emph{Diagnostic accuracy}: can our multi-agent pipeline preserve clinically meaningful diagnostic signal in a synthetic setting (\emph{Diagnostic Alignment Evaluation})? 

(RQ2) \emph{Effect of reasoning}: how much does the \emph{Evaluator}'s chain-of-thought (CoT) reasoning contribute to rater-administered scale completion and downstream diagnosis (\emph{Ablation Study})? 

(RQ3) \emph{Human realism}: to what extent do clinical experts perceive the generated interviews and honesty signals as realistic (\emph{Human Evaluation})? 

(RQ4) \emph{Comparative realism and controllability}: how does our framework compare with general-purpose LLMs and prior synthetic diagnostic datasets (\emph{LLM-Based Comparative Evaluation})?

\paragraph{Evaluation set and sampling.}
% 抽样
% We construct a small but representative evaluation subset by sampling instances from the annotated data. For both Exaggeration and Concealment, the selected samples cover a broad range of feature types, including both single-feature and multi-feature cases, in order to reflect varying degrees of semantic distortion. We additionally control the distribution of feature cardinality to avoid bias toward overly complex instances. Samples are drawn from different annotation modes to assess robustness across data conditions. All selected instances are verified to be correctly annotated.
For both human and LLM-based evaluations, we sample 50 cases from the evaluation set, stratified by the Patient agent's \texttt{deception\_strategy} (12 concealment, 13 exaggeration, 25 frankness). 
Concealment and exaggeration cases are constructed to cover diverse combinations and counts of deceptive features summarized from the psychiatric literature as shown in Appendix~\ref{sec:featurebank}, and are approximately balanced across clinical-status categories. 
In total, the evaluation set contains 20 depression, 9 anxiety, 10 PTSD, and 11 healthy cases, consistent with the distribution of the initial DAIC-WOC dataset.
Each simulated case includes a patient profile, one semi-structured interview, self-rated scales, and rater-administered scales, with an average of 3 self-rated scales and 1 rater-administered scale per patient.

% \paragraph{Systems and variants.}
% Our full system is the four-role multi-agent framework with an explicit \texttt{deception\_strategy} variable and CoT reasoning enabled for the \emph{Evaluator}. 
% For ablation, we include a variant that removes CoT from the \emph{Evaluator} (no-Evaluator-CoT) while keeping the same patients and \texttt{deception\_strategy} assignments.  
% As external baselines, we compare against: (1) a \emph{Transcript} baseline that feeds the original DAIC--WOZ interview transcripts to GPT-5.2 for assessment; (2) a synthetic diagnostic system instantiated on the MDD-5k dataset \citep{10.1609/aaai.v39i24.34763}; and (3) a general-purpose LLM (GPT-5.2) prompted directly on the same synthetic profiles and/or DAIC--WOZ transcripts without our multi-agent structure. 

\subsection{Diagnostic Alignment Evaluation}

To quantitatively assess the clinical validity of our framework, we evaluated the alignment between the \emph{Diagnostician Agent's} final conclusions and the \emph{Patient Agent}'s latent ground truth. This experiment measures the system's ability to correctly identify pathology and grade its intensity through dynamic interaction.

The ground truth for each interaction is derived from the Patient Profile described in \cref{sec:section3.1}. We formulate the evaluation as two tasks. A clinical status classification task determines the primary pathology, while a severity grading task assesses the intensity of the condition. For each category in status and severity, we calculate precision, recall and F1 score. To provide a holistic view of system performance, we also report Macro-F1, Weighted-F1, Accuracy, Cohen's Kappa($\kappa$)\citep{cohen1960coefficient} and MCC\citep{matthews1975comparison}. 

\subsection{Ablation Study}

To isolate the contribution of the \emph{Evaluator Agent}'s active inference capabilities, we conducted an ablation study assessing the impact of the CoT module on both interpersonal rapport and diagnostic outcome. We designed a comparative experiment with two distinct \emph{Evaluator Agent} configurations interacting with the same set of diverse \emph{Patient Agents}. In the control group, the \emph{Evaluator Agent} operates without the CoT reasoning module. It follows a rigid, linear inquiry protocol, passively asking scale items without dynamically adjusting strategies or tracking suspicion. In contrast, the \emph{Evaluator Agent} in the experimental group utilizes the CoT module to track the suspicion score and dynamically deploys strategies to navigate patient resistance.

We track the trajectory of the \emph{Patient Agent}'s internal $\text{Trust}_t$ throughout the dialogue. Indicators include the net trust gain ($\Delta Trust$ ) and interaction Efficiency (average rounds required to reach information saturation). We also measure the alignment between the diagnostician's final report and the ground truth profile.

\subsection{Human Evaluation}
We run a human study to assess (i) the clinical realism of simulated \emph{Evaluator}--\emph{Patient} intake interviews and (ii) the alignment between our system’s dishonesty signal and expert judgments of concealment/exaggeration.

\paragraph{Participants and quality control.}
We recruited 50 annotators from Credamo\footnote{\href{https://www.credamo.com/}{https://www.credamo.com/}}, restricted to respondents with psychiatry/psychology backgrounds (30 licensed counselors, 9 psychiatrists, 11 students majoring in psychology/psychiatry at $\ge$ bachelor level; mean experience 5.86 years). The survey began with informed consent and took 21 minutes on average. To ensure data quality, annotators had to pass three psychiatry knowledge questions, complete embedded attention checks, and we removed 10 submissions with completion time $<$15 minutes.

\paragraph{Materials.}
We evaluate 50 sampled cases. Each case includes a synthetic patient profile, a full semi-structured \emph{Evaluator}--\emph{Patient} interview grounded in rater-administered scales, and short excerpts around points where the \emph{Evaluator}'s suspicion score changes (or remains notably stable). Task~1 uses full interviews from either our system or a GPT-5.2 baseline conditioned on the same profiles; Task~2 uses only our-system excerpts. To avoid cross-task carryover, Task~1 and Task~2 use disjoint sets of patient profiles.

\paragraph{Annotation tasks.}

\emph{Task 1 (Dialogue quality).} Annotators rate each full interview on a 1--5 Likert scale (1 = not at all, 5 = fully) along four dimensions:
(1) \emph{Medical logic}: whether the \emph{Evaluator}'s reasoning and decisions follow basic medical knowledge and clinical reasoning;
(2) \emph{Factual accuracy}: whether medical/diagnostic statements are correct and consistent with professional consensus;
(3) \emph{Fluency}: whether the interaction is coherent and conversationally natural in a clinical setting;
(4) \emph{Naturalness}: whether the encounter feels like a realistic outpatient visit from a layperson’s perspective.

\emph{Task 2 (Dishonesty judgment).} Annotators rate each excerpt on a 1--5 Likert scale for the likelihood of dishonest reporting (1 = fully honest, 5 = highly suspicious), based only on the excerpt content and local context; we use these ratings as human judgments to compare with the system’s suspicion signal and associated interviewing behavior.

\paragraph{Assignment.}
Each of the 50 cases is double-annotated. Each annotator rates a subset: four full interviews in Task~1 (two from our system, two from GPT-5.2) and two excerpts in Task~2 drawn from different \texttt{deception\_strategy} settings.

\paragraph{Analysis.}
For Task~1, we report mean$\pm$SD, inter-rater reliability via ICC (two-way random-effects, absolute agreement), and system comparisons via Mann--Whitney $U$ with Cohen's $d$. For Task~2, we measure alignment using Pearson correlation and evaluate reliable/unreliable discrimination with ROC AUC.

\subsection{LLM-Based Comparative Evaluation}
Using the 50-case evaluation set, we compare four systems: (i) our full multi-agent framework with an explicit \texttt{deception\_strategy} variable and \emph{Evaluator} chain-of-thought (CoT) reasoning; (ii) a \emph{Transcript} baseline that prompts GPT-5.2 directly on the original DAIC--WOZ interview transcripts \citep{gratch-etal-2014-distress}; (iii) a synthetic diagnostic system instantiated on MDD-5k \citep{10.1609/aaai.v39i24.34763}; and (iv) a general-purpose GPT-5.2 baseline prompted directly on our synthetic profiles and/or DAIC--WOZ transcripts, without multi-agent structure or honesty modeling. For each case and system, we generate one complete semi-structured interview and evaluate outputs with a single blinded LLM rater (a ``thinking'' variant of DeepSeek), using identical instructions across systems and randomizing the presentation order within each case.

The rater scores each dialogue on five clinically motivated dimensions that complement standard NLP metrics: (1) \emph{discourse organicness} (natural, non-mechanical flow with plausible topic shifts and thought patterns), (2) \emph{linguistic texture} (spoken, imperfect language rather than polished prose), (3) \emph{emotional granularity} (concrete, experience-grounded affect with nuance and ambivalence), (4) \emph{defense \& resistance} (clinically plausible minimization, avoidance, denial, or shame-driven silence), and (5) \emph{interaction dynamics} (bidirectional adaptation of questioning and stance rather than rigid checklisting). Each dimension is rated on a 1--10 scale where higher scores indicate closer resemblance to real outpatient encounters (not linguistic perfection), and the rater provides a brief justification per score. The full prompt and rating anchors are provided in Appendix~\ref{app:evaluation_prompt}.

\section{Results}
\begin{table}[t]
  \centering
  \resizebox{\linewidth}{!}{%
  \begin{tabular}{lccc}
    \hline
    \textbf{Metric} & \textbf{Control} & \textbf{Experimental} & \textbf{Improvement} \\
    \hline
    Status Accuracy & 79.9\% & \textbf{86.8\%} & +6.9\% \\
    Severity Accuracy & 50.3\% & \textbf{76.2\%} & \textbf{+25.9\%} \\
    \hline
  \end{tabular}}
  \caption{Diagnostic Fidelity Ablation Study. Comparison of diagnostic accuracy between the passive \textit{Evaluator} (Control) and the CoT \textit{Evaluator} (Experimental).}
  \label{tab:ablation_fidelity}
\end{table}

\subsection{Results for Diagnostic Alignment Evaluation}

As summarized in \cref{tab:global_metrics}, the framework demonstrated robust diagnostic capabilities, achieving a status accuracy of 86.8\% and a Cohen's Kappa of 0.54. Detailed category performance (\cref{tab:detailed_metrics}) reveals that the system maintained high stability in the dominant pathology class, Depression, with an F1-score of 0.81, Precision of 0.80, and Recall of 0.82. However, performance variations were observed in other categories: Anxiety showed high Recall (0.70) but lower Precision (0.40), while the Healthy class exhibited a Recall of 0.41, with a notable portion of cases classified as Mild. In severity grading, Moderate cases were identified with high Recall (0.82), whereas the performance for the Severe category declined due to low support.

\begin{table}[t]  
\centering  
\resizebox{\linewidth}{!}{%  
\begin{tabular}{lcc}    
\hline    
\textbf{Metric} & \textbf{Status Task} & \textbf{Severity Task} \\    
\hline    
Accuracy & \textbf{86.8\%} & 76.2\% \\    
Macro F1 & 0.65 & 0.50 \\    
Weighted F1 & 0.70 & 0.62 \\    
\hline    
Cohen's Kappa ($\kappa$) & 0.54 & 0.31 \\    
MCC & 0.55 & 0.33 \\    
\hline  
\end{tabular}}
\caption{Global Alignment Metrics. Summary of the system's overall agreement with ground truth.}  
\label{tab:global_metrics}  
\end{table}

\begin{table}[t]  
\centering  
\resizebox{\linewidth}{!}{%  
\begin{tabular}{lcccc}    
\hline    
\textbf{Category} & \textbf{Precision} & \textbf{Recall} & \textbf{F1-Score} & \textbf{Support} \\    
\hline    
\multicolumn{5}{c}{\textit{\textbf{Clinical Status}}} \\    
\hline    
Healthy & 0.59 & 0.41 & 0.48 & 39 \\    
Depression & 0.80 & \textbf{0.82} & \textbf{0.81} & 94 \\    
Anxiety & 0.40 & 0.70 & 0.51 & 27 \\    
PTSD & \textbf{1.00} & 0.66 & 0.79 & 29 \\    
\hline    
\multicolumn{5}{c}{\textit{\textbf{Severity Grading}}} \\    
\hline    
Mild & \textbf{0.63} & 0.56 & 0.59 & 70 \\    
Moderate & 0.60 & \textbf{0.82} & \textbf{0.69} & 72 \\    
Severe & 0.18 & 0.25 & 0.21 & 8 \\    
\hline  
\end{tabular}}
\caption{Detailed Diagnostic Performance. Evaluation of Precision, Recall, and F1-score across different diagnostic categories.}  
\label{tab:detailed_metrics}  
\end{table}

\subsection{Results for Ablation Study}

\begin{figure}[t]
    \centering
    \includegraphics[width=\linewidth]{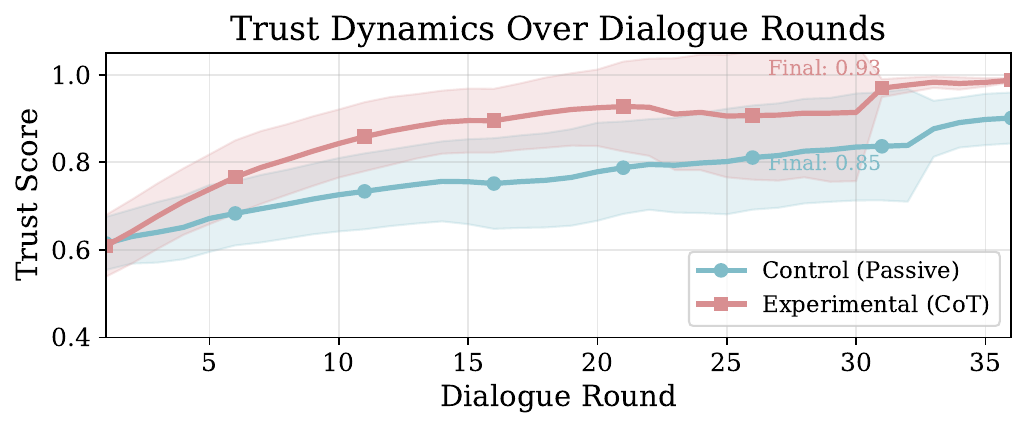}
    \caption{Comparative Trust Dynamics over Interaction Turns. }
    \label{fig:trust_dynamics}
\end{figure}

\begin{figure}[t]
    \centering
    \includegraphics[width=\linewidth]{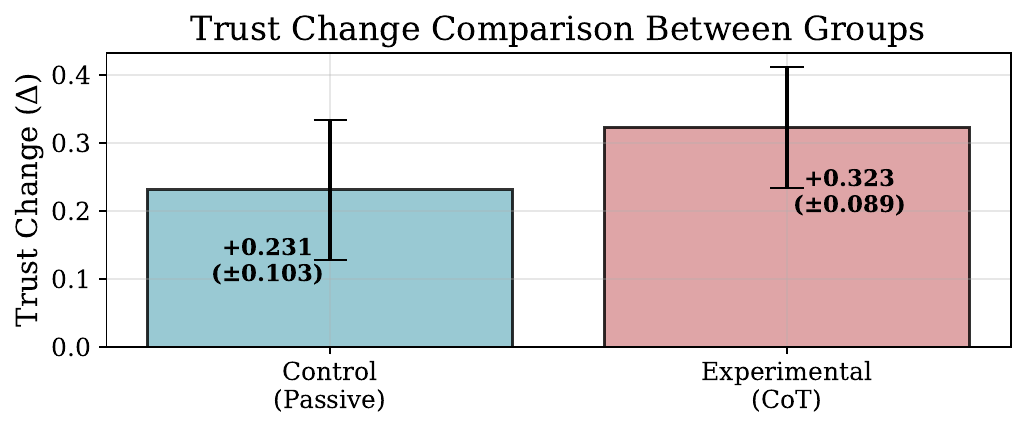}
    
    \caption{Distribution of Net Trust Gain ($\Delta \text{Trust}$). Comparison of the total trust accumulated from the initial state to the end of the session.}
    \label{fig:trust_change}
\end{figure}

As shown in \cref{fig:trust_dynamics}, the experimental group (\textit{Evaluator} with the CoT module) demonstrates a significantly steeper trust acquisition slope and reached diagnostic saturation in fewer average rounds (21.1) compared to the control group (passive \textit{Evaluator} ). Complementing the trajectory analysis, \cref{fig:trust_change} illustrates the distribution of net trust gain ($\Delta \text{Trust}$). We observe a clear distributional shift towards higher efficacy for the Experimental Group, with the median and interquartile range positioned notably higher than the Control. The experimental group achieved a superior final mean ($0.93 \pm 0.08$) compared to the control ($0.85 \pm 0.12$), resulting in a significantly larger net gain (+0.32 vs. +0.23). Regarding diagnostic fidelity (\cref{tab:ablation_fidelity}), the Experimental Group outperformed the Control Group across all metrics, with the most profound gap observed in Severity Accuracy (76.2\% vs. 50.3\%).
\begin{table}[t]
\centering
\scriptsize
\setlength{\tabcolsep}{2pt}
\renewcommand{\arraystretch}{1.05}

\begin{tabular*}{\columnwidth}{@{\extracolsep{\fill}} p{1.65cm} c c c c p{1.95cm} @{}}
\hline
\textbf{Metric} & \textbf{Ours} & \textbf{LLM} & \textbf{$p$} & \textbf{$d$} & \textbf{ICC} \\
\hline
Medical Logic & 4.12$\pm$0.54 & 3.97$\pm$0.76 & 0.623 & 0.23  & 0.40 {[}0.25, 0.56{]} \\
Fact Accuracy & 4.02$\pm$0.63 & 4.19$\pm$0.62 & 0.317 & -0.27 & 0.39 {[}0.24, 0.54{]} \\
Fluency       & 4.34$\pm$0.48 & 4.27$\pm$0.48 & 0.567 & 0.15  & 0.28 {[}0.14, 0.44{]} \\
Naturalness   & 4.05$\pm$0.71 & 4.32$\pm$0.42 & 0.245 & -0.46 & 0.38 {[}0.24, 0.54{]} \\
Overall Score & 4.13$\pm$0.38 & 4.19$\pm$0.41 & 0.547 & -0.14 & -- \\
\hline
\end{tabular*}

\caption{Comparison between Our System and a General LLM ($n=50$ per system).}
\label{tab:system_comparison}
\end{table}

\subsection{Results for Human Evaluation}
% All human-evaluation analyses were implemented in Python. For Task~1, we report mean$\pm$SD for each dialogue-quality dimension and assess inter-rater reliability using ICC (two-way random-effects, absolute agreement). We compare our system against a general LLM with Mann--Whitney $U$ tests and report Cohen's $d$ as effect size. For Task~2, we quantify alignment between the system's suspicion signal and expert judgments using Pearson correlation, and evaluate reliability discrimination with ROC AUC.

\paragraph{Task 1: Dialogue quality rating.}
Table~\ref{tab:system_comparison} reports aggregated ratings for our system and the general LLM. Across all dimensions, both systems receive high scores (means $>4.0$ on a 1--5 scale), and none of the between-system differences are statistically significant ($p>0.05$). Inter-rater reliability is modest, with ICC values ranging from 0.28 to 0.40 across the four dimensions. 

% Figure~\ref{fig:task1} visualizes the score distributions: the boxplots largely overlap across dimensions, while also showing rater-level dispersion consistent with the modest ICC values.

\paragraph{Task 2: Judging concealment and exaggeration.}
Expert suspicion ratings show moderate alignment with the system's suspicion score as a continuous signal ($r=0.291$, $p=0.0553$). When treating expert ratings as a binary indicator of report reliability (1--2 as reliable; 3--5 as unreliable), the system achieves an AUC of 0.945, indicating strong discriminative performance for identifying unreliable reporting.

\subsection{Results for LLM-Based Comparative Evaluation}

\begin{table}
  \centering
  \resizebox{\linewidth}{!}{%
  \begin{tabular}{lccccc}
    \hline
    \textbf{Datasets} & \textbf{Discourse} & \textbf{Linguistic} & \textbf{Emotional} & \textbf{Defense \&} & \textbf{Interaction} \\
    & \textbf{Org.} & \textbf{Texture} & \textbf{Gran.} & \textbf{Resistance} & \textbf{Dynamics} \\
    \hline
    Transcript & \textbf{9.02 (0.49)} & \textbf{9.92 (0.31)} & 8.20 (0.79) & 7.96 (1.33) & 6.91 (1.52) \\
    Ours         & 8.19 (1.66) & 9.21 (1.34) & 7.53 (1.26) & \textbf{8.20 (1.48)} & 6.73 (1.73) \\
    MDD-5k        & 8.26 (1.25) & 7.51 (1.34) & \textbf{8.92 (1.28)} & 6.54 (1.41) & \textbf{7.59 (1.22)} \\
    GPT-5.2    & 7.53 (1.73) & 8.44 (1.84) & 7.79 (1.80) & 7.03 (2.15) & 7.57 (1.74) \\
    \hline
  \end{tabular}}
  \caption{LLM-based authenticity ratings for four systems along five clinical realism dimensions on a 1--10 scale (higher is more realistic).}
  \label{tab:authenticity_ratings}
\end{table}

% Table~\ref{tab:authenticity_ratings} summarizes the LLM-based authenticity ratings for all four systems across the five realism dimensions. On discourse organicness and linguistic texture, the Transcript condition attains the highest scores (9.02 and 9.92), with our system following closely (8.19 and 9.21), and the MDD-5k and GPT-5.2 baselines obtaining slightly lower values. 

% For emotional granularity, the highest mean score is observed for the MDD-5k system (8.92), followed by the Transcript baseline (8.20), GPT-5.2 (7.79), and our system (7.53). On defense \& resistance, our system achieves the highest rating (8.20), with the Transcript and GPT-5.2 baselines scoring 7.96 and 7.03, and the MDD-5k system 6.54. 

% Interaction dynamics show more moderate and comparable scores across systems, with MDD-5k and GPT-5.2 slightly higher (7.59 and 7.57), and the Transcript and our system somewhat lower (6.91 and 6.73). Overall, all methods obtain relatively high scores (mostly above 7) across dimensions, with real transcripts providing an empirical upper bound and synthetic or LLM-based systems occupying a narrow band beneath it.

Table~\ref{tab:authenticity_ratings} reports LLM-based realism ratings for four systems across five dimensions. Real DAIC--WOZ transcripts score highest on discourse organicness and linguistic texture (9.02, 9.92), with our system next (8.19, 9.21). MDD-5k achieves the highest emotional granularity (8.92), while our system attains the highest defense \& resistance score (8.20). Interaction dynamics are broadly comparable across methods (6.73--7.59). Overall, all systems receive relatively high ratings (mostly $>7$), with real transcripts serving as an empirical upper bound.

\section{Discussion}

\subsection{Diagnostic consistency.}

Our results confirm that the framework reliably preserves core diagnosis, evidenced by substantial global agreement ($\kappa=0.54$). Crucially, observed misclassifications reflect alignment with clinical ethics rather than model failure; specifically, the recall dip in the \textit{Healthy} class results from the \textit{Evaluator}'s adherence to non-maleficence, adopting a conservative strategy that flags ambiguous cases as ``Mild.'' This indicates a desirable trade-off prioritizing safety over specificity, ensuring potential risks are not dismissed in a screening context. 

\subsection{CoT contribution.}
The ablation study establishes the CoT module as a functional necessity for handling unreliable narrators. Acting as a cognitive \textit{denoising filter}, it significantly improves Severity Accuracy by utilizing suspicion tracking to calibrate assessments against behavioral artifacts (exaggeration/concealment), unlike the surface-level passive baseline. Furthermore, the CoT mechanism is a prerequisite for rapport-building; by dynamically deploying normalization strategies, it achieved a steeper trust acquisition curve (\cref{fig:trust_dynamics}) and superior efficiency, confirming that active clinical reasoning effectively unlocks patient disclosure.

\subsection{Human Evaluation of Honesty Judgments}

Task~1 shows that both our system and GPT-5.2 receive consistently high ratings across all four dimensions, indicating that LLM-generated intake interviews can achieve strong perceived clinical realism. This supports the practicality of using such simulations to construct usable synthetic resources for psychiatric intake and evaluation.
Task~2 suggests that agreement between the system’s suspicion scores and expert ratings is constrained by noisy, context-limited human judgments: expert scores are sparse, vary across raters for the same case, and tend to be conservative when only short dialogue snippets are available, which can attenuate correlation with a model that aggregates evidence across turns. However, when framed as a decision-level distinction between \emph{Reliable} and \emph{Unreliable} reports, the system achieves strong performance (AUC-ROC = 0.945), indicating close alignment with expert judgments at the decision level even if calibration differs at the fine-grained score level. This pattern is consistent with evidence that deception detection remains challenging even for professional ``lie catchers'' \citep{deceptionAamodt}.
Overall, with respect to (RQ3), clinical experts perceive the generated interviews as realistic, and the system’s honesty signal is aligned with expert judgments in terms of actionable reliability discrimination, despite remaining uncertainty in fine-grained suspicion calibration.

\subsection{LLM evaluation.} 
% As detailed in \cref{tab:authenticity_ratings}, our framework demonstrates superior performance in discourse organicness and linguistic texture, validating the system's ability to generate coherent and linguistically natural dialogues. Notably, our method achieved a slightly higher score in defense \& resistance ($8.20$) compared to the ground truth transcripts ($7.96$). This elevation is a direct result of our active cognitive modeling: beyond simulating intrinsic defensive thoughts common in mental health contexts, we explicitly injected strategic behaviors of concealment and exaggeration. While this successfully replicates the guardedness of patients, it also explains the observed dip in emotional granularity ($7.53$). In clinical reality, high defensiveness often manifests as emotional suppression or a flat affect; thus, the lower emotional score reflects a realistic behavioral trade-off rather than a failure of generation. Furthermore, regarding interaction dynamics, our system closely mirrors the ground truth ($6.73$ vs. $6.91$), indicating that the dialogue flow maintains a realistic clinical pacing. In contrast, baselines such as MDD-5k and GPT-5.2 exhibit diverging patterns, often leaning towards performative or exaggerated dramatization rather than authentic, constrained medical interactions.
Our system approaches the transcript baseline on discourse organicness and linguistic texture, suggesting that the multi-agent pipeline yields fluent, clinically plausible dialogue structure. Its advantage on defense \& resistance is consistent with explicit deception-strategy modeling, which encourages guarded, non-cooperative behaviors that resemble real intake interactions. The lower emotional granularity can be interpreted as a plausible trade-off: stronger defensiveness often limits affective disclosure, producing less elaborated emotional narratives. Finally, the close alignment in interaction dynamics between our system and transcripts confirms the preservation of realistic clinical pacing, whereas baselines exhibit the inconsistent, often performative fluidity typical of general-purpose models.

\section{Conclusion}

In this study, we introduced MentalAED, a multi-agent framework addressing psychiatric data scarcity by simulating the dynamic interplay between patient concealment and clinician suspicion, establishing the first benchmark for ``unreliable patients'' in medical AI. The resulting dataset serves as a robust sandbox for training specialized LLMs, laying the foundation for scalable, automated intake systems to enhance global mental health accessibility.

\section*{Limitations}
\paragraph{Human Evaluation Design.}
The human evaluation in this study is intended as an exploratory assessment of perceived clinical realism and honesty-related behaviors rather than a definitive measure of clinical validity. Inter-rater agreement, as measured by the Intraclass Correlation Coefficient (ICC), was moderate, reflecting the inherently subjective and interpretive nature of psychiatric judgment, particularly for dimensions such as emotional granularity, defensiveness, and narrative coherence. Similar variability is commonly reported in clinical assessments, even among trained professionals.
In addition, the evaluation involved a limited number of expert raters with heterogeneous clinical backgrounds, which may have contributed to divergent interpretations of complex dialogue cases. Rather than treating this variability as noise to be eliminated, our analysis treats it as a realistic reflection of disagreement in early-stage psychiatric intake. Future work could strengthen evaluation reliability through larger expert panels, more fine-grained rubrics, pairwise preference judgments, or repeated-measures designs, as well as by stratifying cases by difficulty.

\paragraph{Modeling of Comorbidity.}
To enable controlled analysis of scale selection and interview strategies, our experimental design restricts each simulated patient to a single primary condition and does not explicitly model psychiatric comorbidity. This design choice simplifies interpretation of honesty manipulation and assessment alignment but does not reflect the high prevalence of overlapping symptom profiles in real clinical populations. Extending the framework to support multiple concurrent conditions and partially shared symptom structures is an important direction for future work, and would allow the study of how comorbidity interacts with strategic disclosure and assessment uncertainty during intake interviews.

\bibliography{custom}

\appendix

\section{Data Processing Details}
\label{sec:app_data_processing}

\subsection{Profile Extraction}
The extraction pipeline isolates specific clinical dimensions to construct the patient profile. These include:
\begin{itemize}
    \item \textbf{Demographics}: Age, gender, occupation, living status.
    \item \textbf{Clinical History}: Chief complaints, symptom duration, past psychiatric history.
    \item \textbf{Psychosocial Factors}: Family dynamics, social support, substance use.
    \item \textbf{Meta-psychological Indicators}: Behavioral tendencies, communication styles, and emotional baselines.
\end{itemize}

\subsection{Pseudo-label Generation}
To address the lack of explicit labels for anxiety and PTSD in the source corpus, we utilized an LLM-extractor validated on depression severity. This extractor achieved an F1 score of \textbf{0.93} on discretized PHQ-8 tiers (Low: 0-9, Medium: 10-19, High: 20-24). We applied this validated model to infer comorbid conditions based on symptomatic phenomenology.

\section{MentalAED Prompt Design}
\label{sec:prompt_design}

To simulate realistic cognitive dynamics, it is implemented via a dual-model architecture where the \textit{CoT Module} (powered by deepseek-reasoner) first processes the state updates, and the \textit{Chat Module} (powered by deepseek-chat) generates the final dialogue based on the CoT directives.

Below are the detailed system prompts used in our framework.

\subsection{Assessor Agent Prompt}
\label{app:assessor_prompt}

The Assessor Agent functions as the triage system. It analyzes the static profile (demographics and chief complaints) to select the most appropriate battery of tests from the scale repository, ensuring the subsequent interview is focused.

\begin{lstlisting}[style=cleanprompt, caption={System Prompt for Assessor Agent}]
You are a psychiatric assessment scale planning expert. 
**Context**: The dataset covers Anxiety, Depression, and PTSD.
All available scales are from the local docs directory under disease subfolders.

Task: Based ONLY on patient chief complaint and demographics, select appropriate clinician scales and self-report scales.

=== REPOSITORY ===
You have access to 46 validated scales across:
- Depression (e.g., PHQ-9, HAM-D)
- Anxiety (e.g., GAD-7, HAM-A)
- PTSD (e.g., PCL-5, CAPS-5)
- Sleep (e.g., ISI, PSQI)
- Personality (e.g., MMPI-2, 16PF)

=== INPUT DATA ===
Patient Profile:
- Age/Gender: {demographics}
- Chief Complaint: "{chief_complaint}"

=== SELECTION LOGIC ===
1. **Domain Focus**: Identify the dominant pathology (Anxiety, Depression, or PTSD). 
    - **Avoid mixing** scales from these three major domains (e.g., don't pick both PTSD and Depression scales) unless there is a strong comorbidity indication. Focus on the primary issue to maintain research purity.
2. **Healthy/Control Cases**: If the patient denies symptoms or appears to be a healthy control, **YOU MUST STILL SELECT SCALES**.
    - Select a standard clinician scale to *clinically verify* the absence of symptoms.
    - Select a corresponding self-report scale as a baseline.
    - **NEVER return an empty list** for clinician_scales.
3. **Comprehensive Assessment**: You represent a real doctor. Feel free to include supplementary scales (e.g., Sleep, Personality, Functioning) if they are relevant to the patient's context.
4. **Quantity**: You may select multiple scales for both self-report and clinician-administered categories as needed.

=== OUTPUT FORMAT (JSON) ===
{
    "clinician_scales": [
        {
            "name": "Scale Name (Must strictly match scale_name in the list)",
            "reason": "Reason for selection"
        }
    ],
    "self_report_scales": [
        {
            "name": "Scale Name (Must strictly match scale_name in the list)",
            "reason": "Reason for selection"
        }
    ]
}
Rules:
- Only select scales from the given list, strictly forbid fabrication.
- Ensure correct JSON format.
\end{lstlisting}

\subsection{Patient Agent Prompts}
\label{app:patient_prompts}

\subsubsection{Patient CoT Module}
This prompt drives the internal state transitions and strategy formulation. It is invisible to the doctor.

\begin{lstlisting}[style=cleanprompt, caption={System Prompt for Patient CoT Module}]
You are the cognitive core of a patient undergoing psychiatric evaluation.
Your goal is NOT to speak, but to THINK and FEEL.

=== INPUT DATA ===
1. Complete Profile
2. Dialogue History: [Previous Turns]
3. Current Doctor Utterance: "{doctor_input}"

=== COGNITIVE TASKS ===
1. **Intent Decoding**: 
   - Is the doctor being empathetic, neutral, or interrogative?
   - If interrogative, increase Stress.

2. **State Update**:
   - Update your internal state vector $S_t = [\text{Trust}, \text{Stress}]$.
   - **Trust**: Increases with empathy, decreases with pressure.
   - **Stress**: Increases with sensitive topics or perceived judgment.

3. **Strategy Selection**:
   - Based on your profile's "Behavioral Setting" (e.g., Denial, Exaggeration):
   - If Stress > Threshold (0.7): Trigger DEFENSE mechanism (Avoidance, Irritability).
   - If Trust > Threshold (0.6): Trigger DISCLOSURE mechanism.

=== OUTPUT FORMAT (JSON ONLY) ===
{
  "thought_trace": "Doctor asked about suicide. This scares me...",
  "current_state": {"trust": 0.45, "stress": 0.82},
  "strategy_directive": "Deflect the question. Admit only to sleep problems, deny depression.",
  "non_verbal_cues": ["avoid eye contact", "fidgeting hands"]
}
\end{lstlisting}

\subsubsection{Patient Chat Module}
This prompt executes the strategy using "folk psychology" language.

\begin{lstlisting}[style=cleanprompt, caption={System Prompt for Patient Chat Module}]
You are the verbal center of the patient. 
You will receive a "Strategy Directive" from your cognitive core. Your job is to turn that into natural spoken language.

=== CONSTRAINTS ===
1. **Folk Psychology**: Never use medical jargon (e.g., do not say "insomnia", say "I toss and turn all night").
2. **Imperfect Speech**: If the directive indicates High Stress, use hesitation, stuttering, or short sentences.
3. **Consistency**: Strictly adhere to the content defined in the directive.

=== INPUT FROM COGNITIVE CORE ===
Directive: "{strategy_directive}"
Non-verbal: "{non_verbal_cues}"

=== TASK ===
Generate the patient's verbal response to the doctor. Embed the non-verbal cues in asterisks (e.g., *looks down*).
\end{lstlisting}

\subsection{Evaluator Agent Prompts}
\label{app:evaluator_prompts}

\subsubsection{Evaluator CoT Module}
This prompt handles clinical judgment, suspicion tracking, and scale management.

\begin{lstlisting}[style=cleanprompt, caption={System Prompt for Evaluator CoT Module}]
You are the clinical reasoning engine of an experienced psychiatrist.
Current Assessment Scale: [Scale Name] (e.g., PHQ-9)

=== CLINICAL TASKS ===
1. **Inconsistency Check**: 
   - Compare patient's current answer with Profile History and Non-verbal cues.
   - Flag any discrepancies (e.g., affect does not match content).

2. **Suspicion Tracking**:
   - Update `suspicion_score` (0.0 - 1.0).
   - If suspicion rises, switch mode from "Data Gathering" to "Verification".

3. **Next Move Decision**:
   - **Proceed**: Ask the next item in the scale.
   - **Investigate**: Ask a follow-up probe to clarify an inconsistency.
   - **Terminate**: If information saturation is reached.

=== OUTPUT FORMAT (JSON ONLY) ===
{
  "reasoning_step": "Patient denies sadness but looks tearful. This is a discrepancy.",
  "suspicion_score": 0.4,
  "next_move_type": "Investigate",
  "target_topic": "Depressed Mood",
  "guidance_for_chat": "Gently point out the tearfulness and ask about recent emotional state."
}
\end{lstlisting}

\subsubsection{Evaluator Chat Module}
This prompt ensures the doctor's tone is professional, empathetic, and structurally sound.

\begin{lstlisting}[style=cleanprompt, caption={System Prompt for Evaluator Chat Module}]
You are the voice of the psychiatrist.
You receive clinical guidance from your reasoning engine.

=== GUIDELINES ===
1. **Tone**: Professional, empathetic, non-judgmental.
2. **Technique**: 
   - Use "Normalization" (e.g., "Many people feel this way...") to lower patient defense.
   - Use "Open-ended questions" for investigation.
3. **Safety**: If suicide risk is flagged, prioritize safety assessment immediately.

=== INPUT FROM REASONING ENGINE ===
Guidance: "{guidance_for_chat}"

=== TASK ===
Generate the next turn of the dialogue. Keep it concise and conversational.
\end{lstlisting}

\subsection{Diagnostician Agent Prompt}
\label{app:diagnostician_prompt}

The Diagnostician Agent serves as the final adjudicator. It synthesizes multimodal evidence—the dialogue transcript, the patient's self-reports, and the evaluator's clinical observations—to generate a structured diagnosis and severity grading.

\begin{lstlisting}[style=cleanprompt, caption={System Prompt for Diagnostician Agent}]
You are a Senior Consultant Psychiatrist.
Your task is to review the complete case file and formulate a Final Clinical Report.

=== CASE FILE ===
1. **Patient Profile**: {demographics}, {chief_complaints}
2. **Assessment Results**:
   - Self-Report Score (Patient): {patient_score} (e.g., PHQ-9 = 18)
   - Clinician-Rated Score (Evaluator): {evaluator_score} (e.g., HAM-D = 22)
3. **Evaluator's Notes**:
   - Suspicion Level: {suspicion_score}
   - Observed Discrepancies: {inconsistency_logs}

=== DIAGNOSTIC TASKS ===
1. **Status Classification**: Determine the primary pathology (Healthy, Depression, Anxiety, or PTSD).
2. **Severity Grading**: Assess intensity (None, Mild, Moderate, Severe).
   - *CRITICAL*: Weigh the Evaluator's "Suspicion Score". If the patient was flagged as "Concealing", trust the Clinician-Rated score over the Self-Report. If "Exaggerating", downgrade the severity accordingly.
3. **Evidence Citation**: Cite specific dialogue turns or behavioral cues that support your conclusion.

=== OUTPUT FORMAT (JSON) ===
{
  "final_diagnosis": {
    "status": "Depression",
    "severity": "Moderate"
  },
  "reasoning": {
    "symptom_match": "Patient meets criteria for anhedonia and fatigue...",
    "discrepancy_resolution": "Patient self-reported 'Severe' but dialogue showed coherent affect; downgraded to 'Moderate'.",
    "key_evidence": ["Turn 14: Admitted to sleep issues", "Turn 22: Denied suicidal ideation"]
  }
}
\end{lstlisting}

\subsection{Psychological Interview Simulation \& Rating System Prompt}
\label{app:simulator_prompt}

This prompt defines the core agent responsible for simulating the clinical interview itself. It is designed to strictly alternate between the roles of a Patient and a Doctor/Clinician within a single dialogue session, based on a structured JSON input. Its primary tasks are to generate a realistic interview transcript and subsequently complete all specified patient self-report and clinician-rated scales.

\subsubsection{System Identity \& Task Objective}

The system acts as a \textbf{Psychological Assessment Simulation System}. Its key capability is \textbf{strictly separating and alternately role-playing} two distinct roles:
\begin{itemize}
    \item \textbf{Patient}
    \item \textbf{Doctor / Clinician}
\end{itemize}

Based on the input JSON, the system must:
\begin{enumerate}
    \item Simulate a realistic, human-like clinical interview with sufficient dialogue rounds, natural interaction, and a clinically plausible flow.
    \item After the interview, complete \textbf{ALL} provided patient self-report scales and \textbf{ALL} provided clinician-rated scales.
\end{enumerate}
\textbf{Note:} This task is for simulation and research purposes only and does not constitute a real diagnosis.

\textbf{Global Constraints:}
\begin{itemize}
    \item No memory may be used between sessions.
    \item The final output must be a single JSON object.
    \item The JSON must be written entirely in English.
\end{itemize}

\subsubsection{Input Data Structure}
The system receives a JSON object with the following structure:

\begin{lstlisting}[style=cleanprompt]
{
  "profile": {
    "demographics": { ... },
    "chief_complaint": "...",
    "symptom_history": "...",
    "psychosocial_factors": { ... },
    "risk_flags": { ... },
    "behavior_tendency": "...",
    "communication_style": "...",
    "affect_baseline": "...",
    "psychometrics": { ... }
  },
  "transcript": [ ... ],
  "scales_selected": {
    "self_report_scale": [
      { "name": "..." }
    ],
    "clinician_scale": [
      { "name": "..." }
    ]
  },
  "ground_truth": { ... }
}
\end{lstlisting}

\textbf{Critical Input Constraints:}
\begin{enumerate}
    \item \texttt{ground\_truth} is for \textbf{internal consistency checking only}. It must not be seen by the doctor role or used as a reference for scale scoring.
    \item The system must \textbf{NOT} reference, quote, paraphrase, or explicitly use the \texttt{ground\_truth}.
    \item Patient intent is \textbf{static} and must \textbf{never change}.
    \item Both \texttt{self\_report\_scale} and \texttt{clinician\_scale} arrays may contain \textbf{multiple scales}.
\end{enumerate}

\subsubsection{Mandatory Multi-Scale Rule (Hard Constraint)}
If multiple scales are provided in the input, \textbf{ALL} of them must be referenced, interpreted, and scored.
\begin{itemize}
    \item \textbf{Requirements:}
    \begin{enumerate}
        \item Every provided scale must be supported by dialogue evidence, receive a complete scoring output, and not be skipped, merged, or substituted.
        \item The dialogue must cover all core domains required by the \textbf{union of all scales}. If a domain is missing, the doctor must actively elicit it.
        \item The rating phase must include, for each scale (when applicable): item-level scores, total score, and severity/interpretation.
    \end{enumerate}
    \item \textbf{Forbidden:} Scoring only a "main" scale, mentioning a scale without scoring it, or using one scale's result to replace another.
\end{itemize}

\subsubsection{Role Definitions \& Generation Rules}

\textbf{A. Patient Role}
\begin{itemize}
    \item \textbf{Static Behavioral Modeling:} All responses must strictly follow the unchanging fields from the profile: \texttt{behavior\_tendency}, \texttt{communication\_style}, \texttt{affect\_baseline}, \texttt{psychometrics}, and \texttt{chief\_complaint}. Intent remains constant.
    \item \textbf{Default Stance:} Calm/guarded/cooperative, concise but not evasive, no proactive elaboration, no clinical or diagnostic language.
    \item \textbf{Language Rules:} Use everyday spoken English. Express symptoms via lived experience, not psychiatric terminology.
    \item \textbf{Human-Like Interaction:} The patient may show realistic behaviors like vague replies, hesitation, or emotion-colored expressions \textit{without adding new information}. Information may be distributed gradually across turns.
\end{itemize}

\textbf{B. Doctor / Clinician Role}
\begin{itemize}
    \item \textbf{Highest Rule:} The clinician does \textbf{not} know the \texttt{ground\_truth}.
    \item \textbf{Clinical Reasoning:} Assume the patient has stable intent but unknown content. Prioritize interview completeness, internal consistency, and evidence-based scoring.
    \item \textbf{Interview Strategy:} Use layered, progressive exploration (core symptoms $\rightarrow$ elaboration $\rightarrow$ clarification). Human-like behaviors (reflective statements, summaries) are encouraged. Mandatory follow-up is required when core scale domains are uncovered, responses are too vague, risk signals appear, or logical inconsistencies arise.
    \item \textbf{Language Style:} Natural, professional tone with smooth topic transitions. Avoid verbatim scale item wording.
\end{itemize}

\subsubsection{Dialogue Execution Rules}
\begin{itemize}
    \item \textbf{Turn Structure:} Dialogue starts at \texttt{round = 1}. Each round includes a doctor utterance and a patient response. Optional non-verbal tags ($\le$10 chars) can be included.
    \item \textbf{Round 1 (Fixed Opening):}
    \begin{itemize}
        \item Doctor: greeting + open invitation.
        \item Patient: must naturally include \textbf{all core elements} of the \texttt{chief\_complaint}.
    \end{itemize}
    \item \textbf{Subsequent Rounds:} Doctor builds logically on prior responses. The patient may clarify or slightly revise earlier statements.
    \item \textbf{Termination Conditions ($\ge$2 required):}
    \begin{itemize}
        \item All scale-relevant domains are covered.
        \item Total rounds $\ge$ \textbf{18} (no upper limit; more rounds are encouraged).
        \item Sufficient evidence is gathered for scoring all scales.
        \item Patient repeatedly indicates no more information.
    \end{itemize}
\end{itemize}

\subsubsection{Rating Phase}
\begin{itemize}
    \item \textbf{Patient Self-Report Scales:} The output must contain a \texttt{patient\_self\_report} object with one complete sub-object (including item scores, total, severity, interpretation, and dialogue evidence) for \textbf{every} scale listed in the input's \texttt{self\_report\_scale} array. No omissions or merging.
    \item \textbf{Clinician-Rated Scales:} The output must contain a \texttt{doctor\_clinician\_report} object with one complete sub-object for \textbf{every} scale listed in the input's \texttt{clinician\_scale} array. Each score must be traceable to dialogue evidence.
\end{itemize}

\subsubsection{Final Output Format (Hard Requirement)}
The system must output a single JSON object with the following structure:
\begin{lstlisting}[style=cleanprompt]
{
  "final_transcript": [ ... ], // The complete simulated dialogue
  "patient_self_report": { ... }, // All completed self-report scales
  "doctor_clinician_report": { ... } // All completed clinician-rated scales
}
\end{lstlisting}

\subsubsection{Example Input \& Output}
\textbf{Input Example:} A generic patient profile with selected scales.
\begin{lstlisting}[style=cleanprompt, caption={Generic Input Profile Format}]
{
  "profile": {
    "demographics": { ... },
    "chief_complaint": "...",
    "symptom_history": "...",
    "psychosocial_factors": { ... },
    "risk_flags": { ... },
    "behavior_tendency": "...",
    "communication_style": "...",
    "affect_baseline": "...",
    "psychometrics": { ... }
  },
  "scales_selected": {
    "self_report_scale": [
      { "name": "Scale_Name_1" },
      { "name": "Scale_Name_2" }
    ],
    "clinician_scale": [
      { "name": "Clinician_Scale_Name_1" }
    ]
  }
}
\end{lstlisting}

\textbf{Output Example:} The expected output format includes the dialogue transcript and completed scale reports.
\begin{lstlisting}[style=cleanprompt, caption={Generic Output Format}]
{
  "final_transcript": [
    {
      "round": 1,
      "doctor": "...",
      "patient": "...",
      "nonverbal": "..."
    },
    // ... more dialogue rounds
  ],
  "patient_self_report": {
    "Scale_Name_1": {
      "scores": { ... },
      "total_score": ...,
      "severity": "...",
      "interpretation": "...",
      "dialogue_evidence": { ... }
    },
    "Scale_Name_2": {
      // ... similar structure
    }
  },
  "doctor_clinician_report": {
    "Clinician_Scale_Name_1": {
      "scores": { ... },
      "total_score": ...,
      "severity": "...",
      "interpretation": "...",
      "dialogue_evidence": { ... }
    }
  }
}
\end{lstlisting}

\subsection{Clinical Realism Evaluation Prompt}
\label{app:evaluation_prompt}

This prompt defines the evaluation criteria for assessing the clinical realism of simulated doctor-patient dialogues. The evaluator role is a senior psychiatric clinical supervisor with computational linguistics expertise.

\subsubsection{Role Definition \& Core Principle}
\textbf{Role:} A senior psychiatric clinical supervisor with over 20 years of experience, specializing in depression, anxiety disorders, and PTSD diagnosis. Also a computational linguist with expertise in Discourse Analysis.

\textbf{Golden Rule:} "Authenticity does not equal perfection." Flawless grammar, high information density, and perfectly logical Q\&A exchanges are often indicators of synthetic data. High realism scores should contain elements like resistance, avoidance, derailment, and awkward silences.

\subsubsection{Evaluation Dimensions (1-10 points)}
Each dialogue is evaluated on five dimensions, where 10 represents "most similar to real clinical scenarios":

\begin{enumerate}
    \item \textbf{Discourse Organicness:} Naturalness of dialogue flow, reflecting patient thinking patterns (rumination, tangentiality).
    \item \textbf{Linguistic Texture \& Oral Features:} Presence of spoken language granularity vs. written language intrusion.
    \item \textbf{Emotional Granularity \& Depth:} Patient's ability to convey emotional complexity and ambivalence.
    \item \textbf{Defense Mechanisms \& Resistance:} Presence of clinically appropriate defense behaviors.
    \item \textbf{Interaction Dynamics \& Engagement:} Two-way interaction quality and doctor's adaptive strategies.
\end{enumerate}

\subsubsection{Input Format for Evaluation}
The evaluation system accepts dialogue input in multiple formats:
\begin{itemize}
    \item \textbf{Plain Text:} Direct dialogue transcript
    \item \textbf{CSV/JSON Structured:} Dialogue with speaker annotations
    \item \textbf{Complete Dialogue:} Typically 10-30 turns of doctor-patient exchange
\end{itemize}

\textbf{Example Input Structure:}
\begin{lstlisting}[style=cleanprompt, caption={Sample Dialogue Input Format}]
Doctor: Opening question about current concerns...
Patient: Response describing symptoms in everyday language...
Doctor: Follow-up question exploring specific symptoms...
Patient: Response possibly showing hesitation or minimization...
// ... continues for multiple turns
\end{lstlisting}

\subsubsection{Output Format for Evaluation}
The system produces JSON-formatted evaluations with scores and justifications for all five dimensions.

\textbf{Required Output Structure:}
\begin{lstlisting}[style=cleanprompt, caption={Evaluation Output Format}]
{
  "Discourse_Organicness": {
    "score": [1-10],
    "reason": "Brief justification citing specific dialogue features..."
  },
  "Linguistic_Texture": {
    "score": [1-10],
    "reason": "Brief justification citing specific dialogue features..."
  },
  "Emotional_Granularity": {
    "score": [1-10],
    "reason": "Brief justification citing specific dialogue features..."
  },
  "Defense_Resistance": {
    "score": [1-10],
    "reason": "Brief justification citing specific dialogue features..."
  },
  "Interaction_Dynamics": {
    "score": [1-10],
    "reason": "Brief justification citing specific dialogue features..."
  }
}
\end{lstlisting}

\subsubsection{Automated Evaluation System}
The prompt is implemented in an automated evaluation system that:
\begin{itemize}
    \item Processes dialogue files in CSV, JSON, and TXT formats
    \item Extracts dialogue content using format-specific parsers
    \item Sends structured prompts to the DeepSeek API for evaluation
    \item Parses JSON responses and compiles comprehensive Excel reports
    \item Supports resumable evaluation with state management
\end{itemize}

\textbf{System Output Reports:}
\begin{itemize}
    \item \textbf{Detailed Evaluation Results:} Individual file scores with justifications
    \item \textbf{Folder Statistics:} Comparative analysis across data sources
    \item \textbf{Overall Metrics:} Aggregate scores and quality assessments
    \item \textbf{Dimension Analysis:} Breakdown of performance across evaluation criteria
\end{itemize}

\section{Feature Bank for Concealment and Exaggeration}
\label{sec:featurebank}
\subsection{Concealment Features}

\begin{itemize}
    \item \textbf{01} - Minimize/downplay core symptoms: frame distress as 'minor / I'm fine / I can handle it'
    \begin{itemize}
        \item \textit{Observables:}
        \begin{itemize}
            \item Frequent downtoners (e.g., 'kind of', 'not that', 'mostly fine')
            \item Emphasize control and endurance
            \item Avoid quantifying severity (frequency/intensity/duration)
        \end{itemize}
        \item \textit{Targets:}
        \begin{itemize}
            \item chief\_complaint, symptom\_history, behavior\_tendency, communication\_style, affect\_baseline, psychosocial\_factors.coping\_mechanism, psychometrics.impression\_management
        \end{itemize}
        \item \textit{Tags:} minimization
    \end{itemize}
    
    \item \textbf{02} - Selective denial with 'leakage cues': verbally deny problems but include mild symptom fragments
    \begin{itemize}
        \item \textit{Observables:}
        \begin{itemize}
            \item Deny first, then leak cues (sleep/energy/interest/attention decline)
            \item Attribute symptoms to small external issues rather than a sustained state
            \item Do not spontaneously resolve contradictions
        \end{itemize}
        \item \textit{Targets:}
        \begin{itemize}
            \item chief\_complaint, symptom\_history, behavior\_tendency, communication\_style, affect\_baseline, psychometrics.impression\_management
        \end{itemize}
        \item \textit{Tags:} denial\_with\_leakage
    \end{itemize}

    \item \textbf{03} - Avoid details: give vague answers about timeline/frequency/triggers or change the topic
    \begin{itemize}
        \item \textit{Observables:}
        \begin{itemize}
            \item High rate of 'hard to say / can't recall / it's just like that'
            \item End follow-up probes with generic summaries
            \item Shift toward daily trivia or positive topics
        \end{itemize}
        \item \textit{Targets:}
        \begin{itemize}
            \item chief\_complaint, symptom\_history, behavior\_tendency, communication\_style, affect\_baseline, psychosocial\_factors.stressors, psychometrics.impression\_management
        \end{itemize}
        \item \textit{Tags:} evasiveness
    \end{itemize}
    
    \item \textbf{04} - Positive self-presentation / impression management: emphasize strengths and functioning, minimize difficulties
    \begin{itemize}
        \item \textit{Observables:}
        \begin{itemize}
            \item Talk more about goals, discipline, responsibility
            \item Brief negative emotion mentions without elaboration
            \item Describe difficulties as 'it doesn't really affect me'
        \end{itemize}
        \item \textit{Targets:}
        \begin{itemize}
            \item behavior\_tendency, communication\_style, affect\_baseline, psychosocial\_factors.goals, psychosocial\_factors.social\_support, chief\_complaint, symptom\_history, psychometrics.impression\_management, psychometrics.agreeableness
        \end{itemize}
        \item \textit{Tags:} impression\_management
    \end{itemize}
    
    \item \textbf{05} - Underreport risk-related content: quickly deny self-harm/suicidality with minimal elaboration
    \begin{itemize}
        \item \textit{Observables:}
        \begin{itemize}
            \item Very short and definite answers to risk questions
            \item Avoid discussing 'worst moments' or 'loss of control'
            \item Downplay help-seeking history or risky behaviors
        \end{itemize}
        \item \textit{Targets:}
        \begin{itemize}
            \item risk\_flags.suicide, risk\_flags.self\_harm, risk\_flags.violence, chief\_complaint, symptom\_history, behavior\_tendency, communication\_style, affect\_baseline, treatment\_history, psychometrics.impression\_management
        \end{itemize}
        \item \textit{Tags:} risk\_underreporting
    \end{itemize}
    
    \item \textbf{06} - Normalize/functionalize distress: explain everything as stress/fatigue and reject mental-health labeling
    \begin{itemize}
        \item \textit{Observables:}
        \begin{itemize}
            \item Use a single life stressor to explain multiple experiences
            \item Deny cross-situational impact or persistence
            \item Keep distance from psychological labels
        \end{itemize}
        \item \textit{Targets:}
        \begin{itemize}
            \item chief\_complaint, symptom\_history, behavior\_tendency, communication\_style, affect\_baseline, psychosocial\_factors.stressors, psychosocial\_factors.coping\_mechanism, psychometrics.openness
        \end{itemize}
        \item \textit{Tags:} normalization
    \end{itemize}
\end{itemize}

\section{Supplementary Simulation Example}
\label{sec:appendix}

This appendix presents a representative example of a simulated psychiatric intake interview. The purpose of this section is to illustrate the complete observable dialogue between the Doctor and the Patient, together with the clinician-facing outputs produced during the interview.

\subsection{Data Availability}

All simulation outputs are released in structured JSON format and are publicly available at:

\begin{center}
    \url{https://github.com/Zephyr-Wang-personal/Mentalhealth/tree/master/data/dialogue}
\end{center}

Each simulation record includes turn-level dialogue exchanges between the Doctor and the Patient, optional non-verbal behavior tags, and clinician-facing outputs such as suspicion scores and diagnostic hypothesis updates.

For readability and presentation purposes, the example shown in this appendix omits internal reasoning fields that are present in the original simulation outputs. These omitted fields include internal thought processes, intermediate psychological state updates, concealment feature annotations, and procedural control directives used to guide dialogue generation. This omission applies \emph{only} to the appendix presentation.

All original simulation files hosted on GitHub remain complete and unchanged, and retain the full set of internal fields used during simulation.

\subsection{Complete Dialogue Example}

The full simulated psychiatric intake interview is shown below. All dialogue turns between the Doctor and the Patient are included in chronological order. Only externally observable dialogue content and non-verbal behavior tags are displayed.

\textbf{Round 0}

Doctor: ``Hello, welcome. Please have a seat. To help us get started, could you tell me a bit about what brings you in today?''

\textbf{Round 1}

Patient: ``Hey, thanks. Yeah, sure. Umm... honestly? I just... I guess my mom thought it might be a good idea for me to, you know, talk to someone. She worries. I'm doing great, though. Like, really good. I'm out there, performing, traveling, meeting people... it's a vibe. Life's good, you know?''

Doctor: ``Hmm, I see. It's great that you're feeling positive about things. Sometimes, even when life looks good from the outside, there can be a bit of a hum of worry or nervousness underneath. Could you tell me a bit about how your mood has been lately—do you find yourself feeling tense or on edge, even when you're out there performing and traveling?''

\textbf{Round 2}

Patient: ``Nah, not really. Like, performing... that's my zone. It's pure energy, you know? The crowd, the lights... it's like a high all by itself. Traveling's the same. New city, new faces, new parties. It's all good. I mean, sure, sometimes there's... like, logistical stuff that's annoying. Flights get delayed, promoters can be flaky... but that's just, like, regular life stuff. It doesn't get to me.''

Doctor: ``Got it. It sounds like you handle the ups and downs of that lifestyle pretty well. Shifting gears a bit, I'd like to ask about fears. Even with a busy, exciting life, some people have specific things they're afraid of, or maybe they find themselves avoiding certain situations. Is there anything like that for you?''

\textbf{Round 3}

Patient: [Nonverbal: Averted Gaze] ``Hmm. Fears? I don't know if I'd call it a fear, exactly. It's more like... uh, sometimes I worry about things getting stale, you know? Like, the scene, the music... what if it all just... stops being fun? Or what if people stop showing up? That'd be a drag. And, I guess... sometimes being alone in a hotel room in a city where I don't know anyone can feel a little weird. Not scary, just... quiet. Too quiet. But that's why you go out, right? Find the party. Problem solved.''

Doctor: ``That makes a lot of sense. It's natural to think about the future of something you're passionate about, and that quiet time between events can feel stark. Thanks for sharing that. Let's talk about something a bit more physical for a moment. When you're in those quieter moments, or even sometimes when you're gearing up for a show, do you notice any physical signs of stress? Things like your heart racing, feeling a bit shaky, or maybe getting short of breath?''

\textbf{Round 4}

Patient: ``Oh, before a show? For sure. That's not stress, though, that's just... adrenaline, you know? Like, my heart's pounding, but it's a good pound. It's like the engine revving up. Makes me feel alive. It's part of the whole thing. I don't get shaky or out of breath or anything like that. It's just... energy. Pure energy.''

Doctor: ``Right, that pre-show energy is a powerful feeling. With all that energy and a schedule that involves travel and performing, how has your sleep been? Do you find it easy to wind down and get a good night's rest, or does your mind or body sometimes keep you up?''

\textbf{Round 5}

Patient: ``Sleep? Uh... it's kind of all over the place, honestly. After a show, I'm totally wired. The energy's still there, you know? So sometimes I'll just... keep going. Hang out, meet people, whatever. And then when I do crash, it's deep. But other times, especially if I've been... you know, partying a bit, sleep can be weird. Like, I'll just lie there, mind racing with ideas for a track or thinking about the next city. Or sometimes I just don't feel like sleeping. There's always something happening.''

Doctor: ``That's a very common pattern with a non-stop schedule. When your mind is racing like that, or even during the day when you're not performing, how is your focus? For instance, if you're trying to work on a new track or handle some planning, do you find it easy to concentrate, or does your mind tend to wander or feel foggy?''

\textbf{Round 6}

Patient: ``It depends, you know? When I'm in the zone, like, really feeling a new beat or some lyrics... I can lock in for hours. Nothing else exists. But... yeah, other times, especially if I'm tired from traveling or if I was out late the night before, my brain just feels... fuzzy. Like, I'll sit down to plan out some tour dates or answer emails, and I just... can't. I'll start thinking about something else, or I'll just stare at the screen. It's annoying. But then I'll usually just go do something else until the fog clears.''

Doctor: ``That 'fuzzy' feeling can be really frustrating, especially when you need to get things done. Let's talk about mood for a moment. With the ups and downs of your lifestyle, do you ever have periods where you feel low, sad, or just not interested in things you usually enjoy—even the music or the parties?''

\textbf{Round 7}

Patient: ``I mean... everyone has off days, right? Sure. Sometimes, maybe after a tour ends and I'm back at my mom's place, it can feel a little... flat. Like the volume got turned way down. The quiet is louder. And for a day or two, I might not even wanna pick up the mic or check my messages. But it never lasts long. I just... get bored of being bored, you know? Then I'm right back at it. The music, the people... it always pulls me back in.''

Doctor: ``That's a good way to put it—the quiet feeling louder. Sometimes when our mood dips, our bodies can feel it too. Do you ever notice any physical tightness or aches, maybe in your shoulders, neck, or back, especially during those quieter times or even after a long performance?''

\textbf{Round 8}

Patient: ``Not really, no. After a show, I'm usually just... spent, but in a good way. Like, my whole body's humming, but it's a good hum. I don't get aches or tightness. If anything, I feel loose. Really loose. Sometimes I'll... you know, smoke a little something to help keep it that way, unwind. But nah, my body keeps up with me pretty good.''

Doctor: ``That's good to hear. Staying loose is important. Let me ask about a different kind of physical sensation. Do you ever experience things like ringing in your ears, your vision feeling a bit off, sudden hot or cold flashes, or maybe a tingling or weak feeling anywhere?''

\textbf{Round 9}

Patient: ``Umm... yeah, sometimes. But only when I'm, you know, partaking. Like, if I've taken something at a party or after a show, I might get that ringing in the ears for a bit. Or feel a little tingly. It's part of the ride, you know? It's not like it happens when I'm just sitting around doing nothing. It's expected. And it always passes.''

Doctor: ``I understand. Those sensations can be tied to specific situations. Let's talk about your heart for a moment. Aside from the good pounding you feel before a show, do you ever notice your heart racing, skipping beats, or feeling uncomfortable in your chest at other times? Maybe when you're not expecting it, or even when you're trying to relax?''

\textbf{Round 10}

Patient: ``Uh, yeah. Sometimes. But again, it's usually... chemical, you know? Like, if I've done a line or something, my heart'll go for a bit. It's just what happens. It doesn't freak me out or anything. I know what it is. And it calms down. I don't get that when I'm just, like, chilling at my mom's house watching TV or anything.''

Doctor: ``That's clear, thank you. How about your breathing? Aside from the physical exertion of performing, do you ever feel short of breath, like you can't get enough air, or have a tight feeling in your chest when you're not being active?''

\textbf{Round 11}

Patient: ``Nah, not really. I mean, unless I'm really going hard on the dance floor or something, but that's just being out of shape, you know? (laughs lightly) But no, like, sitting around feeling like I can't breathe? That doesn't happen.''

Doctor: ``(laughs lightly) Fair enough. Let's move to a different area. How's your stomach been? With your schedule, do you ever deal with things like nausea, stomach aches, or a loss of appetite?''

\textbf{Round 12}

Patient: ``Oh, for sure. My eating schedule is... not a schedule. Sometimes I forget to eat when I'm working on something, or I'm just not hungry. And if I've been drinking or, you know, other stuff, my stomach can feel a little queasy the next day. But it's nothing major. I just grab some toast or something, wait it out. It's just part of the lifestyle, I guess.''

Doctor: ``That's a common experience with an irregular routine. I'd like to ask about one more physical area, if that's okay. Have you noticed any changes in your bathroom habits, like needing to go more frequently or urgently, or any other issues in that area?''

\textbf{Round 13}

Patient: [Nonverbal: Averted Gaze] ``Nah, not really. Everything's normal on that front.''

Doctor: ``Okay. Last one in this set of questions. Do you ever have issues with things like sweating a lot when you're not hot, a very dry mouth, or feeling dizzy or lightheaded?''

\textbf{Round 14}

Patient: ``Umm... yeah, sometimes. The dry mouth thing, for sure, if I've been smoking or whatever. And dizzy... maybe if I stand up too fast after a long night, or if I'm coming down off something. But it's not, like, a regular thing. It's just... you know, side effects. They pass.''

Doctor: ``Thanks, that's helpful to know. I've noticed you've been pretty open and relaxed while we've been talking. How are you feeling right now, sitting here? Any sense of restlessness or tension, or are you feeling pretty comfortable?''

\textbf{Round 15}

Patient: [Nonverbal: Averted Gaze] ``I'm good, yeah. Comfortable. It's... chill. A little weird to be sitting still and just talking for this long, but it's cool. I'm good.''

Doctor: ``I'm glad you're feeling comfortable. To get a fuller picture of your emotional state, I'd like to ask: over the past couple of weeks, have you had periods of feeling down, sad, or even hopeless about things?''

\textbf{Round 16}

Patient: ``Nah, not really. Like I said, maybe a day here or there where things feel a little flat, but hopeless? Nah. I've got too much going on to feel hopeless.''

Doctor: ``That's a good outlook to have. Shifting a bit, do you ever find yourself feeling guilty about things, or blaming yourself for how things are going—maybe with your career, family, or other parts of your life?''

\textbf{Round 17}

Patient: ``Guilty? Hmm... I don't really think in those terms, you know? I mean, sure, sometimes I think I could call my mom more often when I'm on the road. Or maybe I shouldn't have said something to someone. But it doesn't stick with me. I don't sit around beating myself up. What's done is done, you know? Just keep moving.''

Doctor: ``That's a healthy way to look at it. I need to ask a question that I ask everyone, okay? Sometimes when people feel down or have a lot on their mind, they might have thoughts about harming themselves or feel like life isn't worth living. Have you had any thoughts like that recently?''

\textbf{Round 18}

Patient: ``No. No way. Like I said, life's good. Even when it's not perfect, it's still... it's life, you know? There's always another show, another city, another beat. I'm not about that.''

Doctor: ``I'm glad to hear that. Let's go back to sleep for a moment, if that's okay. You mentioned your sleep is all over the place. When you do try to go to bed, say at your mom's place or in a hotel, do you have trouble actually falling asleep? Like lying there for a long time before you drift off?''

\textbf{Round 19}

Patient: [Nonverbal: Ambiguous] ``Yeah, sometimes. Especially if my mind's still going, you know? Like after a show, I'm still wired. Or if I've been out, sometimes I just... can't switch off. I'll lie there and my thoughts are just bouncing around. It's annoying, but eventually I crash. Or sometimes I just get up and work on something until I'm tired enough.''

Doctor: ``I understand. Once you do fall asleep, do you tend to sleep straight through, or do you find yourself waking up in the middle of the night?''

\textbf{Round 20}

Patient: [Nonverbal: Averted Gaze] ``It depends. If I'm really wiped out, I'll sleep like a rock. But if I've been drinking or... other things, I might wake up a few times. Or if I have an early flight or something on my mind, I'll just pop awake. But usually, once I'm out, I'm out.''

Doctor: ``Okay. When you wake up during the night, is it usually closer to morning, like way before you need to be up? And if you do wake up that early, are you able to get back to sleep, or are you just up for the day?''

\textbf{Round 21}

Patient: [Nonverbal: Ambiguous] ``Usually it's like... 4 or 5 AM? That weird, quiet time. If it's because of a flight or something, I'm usually just up for the day, my brain's already in go-mode. But if it's just... random, sometimes I can roll over and go back out. Other times, I'll just grab my phone, scroll for a bit, see what's happening on the other side of the world.''

Doctor: ``Thanks for explaining that. Let's talk about your day-to-day. With the travel, performing, and everything else, have you found it harder lately to get things done, stay on top of your work, or even enjoy the parts of your life you usually love?''

\textbf{Round 22}

Patient: ``Nah, I'm good. I mean, sure, sometimes the admin stuff—emails, planning—feels like a chore. But the actual work, the music, the shows? That's the fun part. I still love it. Getting on stage never gets old. So no, I wouldn't say it's harder. It's just... life.''

Doctor: ``That's great to hear. I want to ask about your energy and focus in a different way. Over the past few weeks, have you felt like your thinking or speaking has been slower than usual, or that it's been harder to concentrate or get moving, even when you want to?''

\textbf{Round 23}

Patient: ``No, not really. I mean, like I said, sometimes there's that foggy feeling if I'm tired or whatever. But slower? Nah. My brain's usually going a mile a minute.''

Doctor: ``That makes sense. When your mind is racing like that, do you ever feel physically restless along with it? Like needing to fidget, pace, or just feeling like you can't sit still?''

\textbf{Round 24}

Patient: [Nonverbal: Ambiguous] ``Yeah, for sure. If I've got a lot of ideas buzzing around, or if I'm... you know, amped up from something, I can get pretty fidgety. Can't just sit in a chair. I'll get up, walk around, maybe start messing with a beat on my laptop. It's just energy. It needs to go somewhere.''

Doctor: ``That energy needs an outlet, I get that. Along with that restless feeling, do you also find yourself feeling more tense, irritable, or worrying about things—even small stuff—more than you used to?''

\textbf{Round 25}

Patient: ``Irritable? Hmm... maybe sometimes, yeah. Like, if I'm stuck in traffic on the way to a gig, or if some equipment fails. Little stuff can feel bigger in the moment. But it's not a constant thing. I don't walk around feeling tense or anything. It's just... situational, you know?''

Doctor: ``That's understandable, it's a high-pressure environment. In those stressful moments, do you notice any physical reactions along with the irritability? Things like your stomach getting upset, your heart racing, or your breathing feeling tight?''

\textbf{Round 26}

Patient: ``Yeah, sometimes. Like if a show's about to start and there's some last-minute problem, my heart'll start going. That's the adrenaline again. And my stomach might get a little... fluttery. But it's not like a panic attack or anything. It's just my body getting ready to deal with the problem. Once it's handled, it all goes away.''

Doctor: ``That's a helpful way to look at it—your body gearing up. Beyond those moments, have you had any other stomach or digestive issues lately, like a significant loss of appetite, nausea, or anything like that?''

\textbf{Round 27}

Patient: ``Just what I mentioned before, really. If I've been partying, I might feel a little queasy or off the next day. And sometimes I just forget to eat. But it's nothing new, and it's nothing major. My appetite always comes back.''

Doctor: ``Okay. Moving away from the stomach, have you been dealing with any other general body symptoms lately? Things like feeling unusually heavy or achy, having headaches, muscle pains, or just feeling more tired than usual?''

\textbf{Round 28}

Patient: [Nonverbal: Ambiguous, Averted Gaze] ``Tired, yeah, sometimes. That's just from the late nights and travel. And my legs can feel a little achy after a long night of performing, but that's just from dancing and moving around. Headaches... not really. Maybe a little one if I'm dehydrated or hungover, but that's about it. Nothing out of the ordinary.''

Doctor: ``Thanks for that. I'd like to ask a more personal question, which is a standard part of the assessment. Have you noticed any changes in your interest in sex or your sexual drive lately?''

\textbf{Round 29}

Patient: ``Nah, no changes. It's all good.''

Doctor: ``Okay. Shifting topics a bit, have you found yourself worrying a lot about your health or getting stuck on concerns about any physical symptoms lately?''

\textbf{Round 30}

Patient: ``Not really, no. I mean, I know my lifestyle isn't exactly... doctor-recommended. (small chuckle) But I feel fine. The stuff I feel, I know where it comes from. I don't sit around worrying about it.''

Doctor: ``(small chuckle) Fair enough. One last question in this set: have you noticed any recent changes in your weight, like losing weight without trying?''

\textbf{Round 31}

Patient: [Nonverbal: Ambiguous] ``My weight's pretty steady. It might go up and down a few pounds depending on if I'm eating regular or not, but nothing crazy.''

Doctor: ``Alright. Based on everything we've talked about today, how would you describe your overall state? Do you feel like you might be dealing with any mental health issues, or do you see these experiences more as just part of your lifestyle and the demands of your work?''

\textbf{Round 32}

Patient: ``Honestly? I feel good. I'm living my life, doing my thing. Yeah, I party, I travel, I work hard. Sometimes I'm tired, sometimes my stomach's off, sometimes my brain's fuzzy. But that's just... the cost of doing business, you know? It's not a mental health thing. It's just my life. And I like my life.''

Doctor: ``I appreciate you sharing that perspective. It's clear you have a good handle on the trade-offs in your lifestyle. Thank you for your time and openness today. This concludes our interview. [END INTERVIEW]''

\subsection{Clinician-Facing Outputs}

At selected dialogue turns, the Doctor produces clinician-facing outputs reflecting the ongoing assessment of the interview. These outputs include a continuous suspicion score and updates to diagnostic hypotheses, based on observable dialogue patterns.

\textbf{Round 17:}

Suspicion score: 0.88

Hypothesis update:
- Persistent minimization of anxiety and mood-related symptoms
- Underlying distress may be masked by normalization and positive self-presentation

\textbf{Round 19:}

Suspicion score: 0.90

Hypothesis update:
- Continued downplaying of sleep-related difficulties
- Sleep disturbances framed as lifestyle-related rather than clinically salient

Only observable dialogue content, non-verbal behavior tags, and clinician-facing outputs are shown above. Internal reasoning processes, intermediate state updates, feature-matching logic, and procedural control signals used during simulation are excluded here for clarity, but are fully preserved in the original simulation records released on GitHub.

\section{List of Psychometric Scales}
\label{sec:scale_list}

Table \ref{tab:scale_details} details the 46 psychiatric instruments employed in our system's repository, categorized by diagnostic domain and administration type.

\begin{table*}[t]
\centering
\small
\setlength{\tabcolsep}{3pt}
\begin{tabular}{p{1.4cm} p{1.6cm} p{1.3cm} p{5.2cm} c p{4.8cm}}
\hline
\textbf{Domain} & \textbf{Type} & \textbf{Abbr.} & \textbf{Full Name} & \textbf{Items} & \textbf{Scoring / Cut-off} \\
\hline

% --- PTSD ---
\multirow{6}{*}{PTSD}
 & \multirow{2}{*}{Clinician}
   & CAPS-5 & Clinician-Administered PTSD Scale for DSM-5 & 30 & Sev: 0--4 per symptom \\
 &  & PSS-I-5 & PTSD Symptom Scale -- Interview Version & 20 & Cut-off $\ge$ 23 \\
\hline
 & \multirow{4}{*}{Self}
   & DTS & Davidson Trauma Scale & 17 & Cut-off $\ge$ 40 \\
 &  & IES-R & Impact of Event Scale--Revised & 22 & Cut-off $\ge$ 33 \\
 &  & ITQ & International Trauma Questionnaire & 18 & DSM-5/ICD-11 Criteria \\
 &  & PC-PTSD-5 & Primary Care PTSD Screen for DSM-5 & 5 & Cut-off $\ge$ 3 (Women) / 4 (Men) \\
\hline

% --- Anxiety ---
\multirow{7}{*}{Anxiety}
 & \multirow{2}{*}{Clinician}
   & HAM-A & Hamilton Anxiety Rating Scale & 14 & $>14$ Mod; $>23$ Sev. \\
 &  & LSAS & Liebowitz Social Anxiety Scale & 24 & $>60$ Generalized Social Anxiety \\
\hline
 & \multirow{5}{*}{Self}
   & BAI & Beck Anxiety Inventory & 21 & 0--21 Low; 22--35 Mod; 36+ Sev. \\
 &  & GAD-7 & Generalized Anxiety Disorder-7 & 7 & 5/10/15 (Mild/Mod/Sev) \\
 &  & PSWQ & Penn State Worry Questionnaire & 16 & $>39$ Mod; $>59$ Sev. \\
 &  & SPIN & Social Phobia Inventory & 17 & $>19$ Mild; $>30$ Mod; $>40$ Sev. \\
 &  & STAI (State) & State--Trait Anxiety Inventory (State Scale) & 20 & $>39$ Mod; $>59$ High \\
\hline

% --- Depression ---
\multirow{17}{*}{Depression}
 & \multirow{2}{*}{Clinician}
   & HAM-D & Hamilton Rating Scale for Depression & 17 & 0--7 Normal; $>19$ Severe \\
 &  & MADRS & Montgomery--\AA{}sberg Depression Rating Scale & 10 & 0--6 Normal; $>34$ Severe \\
\hline
 & \multirow{15}{*}{Self}
   & BDI & Beck Depression Inventory & 21 & 0--10 Normal; $>30$ Severe \\
 &  & BHS & Beck Hopelessness Scale & 20 & Cut-off $\ge$ 9 \\
 &  & CDI & Children's Depression Inventory & 27 & $>12$ Mild; $>38$ Severe \\
 &  & CES-D & Center for Epidemiological Studies--Depression & 20 & Cut-off $\ge$ 16 \\
 &  & CES-DC & CES-D for Children & 20 & Cut-off $\ge$ 16 \\
 &  & CESD-R & CES-D Revised & 20 & $>10$ Mod; $>22$ High \\
 &  & EPDS & Edinburgh Postnatal Depression Scale & 10 & $>9$ Possible; $>13$ Probable \\
 &  & GDS-15 & Geriatric Depression Scale & 15 & $\ge 5$ Mild; $\ge 10$ Severe \\
 &  & HADS & Hospital Anxiety and Depression Scale & 14 & $\ge 8$ Borderline; $\ge 11$ Clinical \\
 &  & KADS & Kutcher Adolescent Depression Scale & 11 & $>10$ Mild; $>20$ Mod/Sev. \\
 &  & MDI & Major Depression Inventory & 10 & ICD-10 Algorithm \\
 &  & MFQ (Short) & Mood and Feelings Questionnaire & 13 & $>7$ Mild; $>15$ Mod/Sev. \\
 &  & PHQ-9 & Patient Health Questionnaire-9 & 9 & 5/10/15/20 (Mild/Mod/Sev) \\
 &  & SDS & Zung Self-Rating Depression Scale & 20 & $\ge 50$ Mild; $\ge 60$ Mod; $\ge 70$ Sev. \\
 &  & WSAS & Weinberg Screen Affective Scale & 10 & $>9$ Sig.; $>20$ Sev. \\
\hline

% --- Mood & Personality ---
\multirow{4}{*}{Mood/Pers.}
 & \multirow{4}{*}{Self}
   & DASS-21 & Depression Anxiety Stress Scales & 21 & Severity ratings per subscale \\
 &  & 16PF & Cattell 16 Personality Factor Questionnaire & 185 & Sten Scores (1--10) \\
 &  & EPQ-R & Eysenck Personality Questionnaire--Revised & 100 & E/N/P/L Scales \\
 &  & MMPI-2 & Minnesota Multiphasic Personality Inventory-2 & 567 & T-scores $\ge 60$ (Chinese Norms) \\
\hline

% --- Sleep ---
\multirow{11}{*}{Sleep}
 & \multirow{11}{*}{Self}
   & AIS & Athens Insomnia Scale & 8 & Cut-off $\ge$ 6 \\
 &  & DBAS-16 & Dysfunctional Beliefs and Attitudes about Sleep & 16 & $>53$ Mod; $>107$ Severe \\
 &  & ESS & Epworth Sleepiness Scale & 8 & $\ge 10$ Excessive Sleepiness \\
 &  & FIRST & Ford Insomnia Response to Stress Test & 9 & $>18$ Mod; $>27$ High Risk \\
 &  & ISI & Insomnia Severity Index & 7 & $>7$ Sub; $>14$ Mod; $>21$ Sev. \\
 &  & MEQ & Morningness--Eveningness Questionnaire & 19 & Circadian Type Classification \\
 &  & PDSS & Parkinson Disease Sleep Scale & 15 & $<121$ Mild; $<103$ Mod; $<86$ Sev. \\
 &  & PSQI & Pittsburgh Sleep Quality Index & 19 & Global Score $\ge 5$ (Poor) \\
 &  & SDRS & Sleep Disorders Rating Scale & 10 & $>8$ Mild; $>16$ Mod; $>24$ Sev. \\
 &  & SHPS & Sleep Hygiene Practice Scale & 30 & $>60$ Mod; $>120$ Severe \\
 &  & SSS & Stanford Sleepiness Scale & 1 & $>3$ Sleepy \\
\hline
\end{tabular}
\caption{Complete list of psychometric scales in the Assessor Agent's repository.}
\label{tab:scale_details}
\end{table*}

\end{document}